\documentstyle[a4p,12pt,epsfig,cite]{article}

\begin{document}
\begin{titlepage}
\begin{center}{\large   EUROPEAN LABORATORY FOR PARTICLE PHYSICS
}\end{center}\bigskip
\begin{flushright}
CERN-PPE/97-063 \\
June 5, 1997  
\end{flushright}
\begin{center}
{\huge\bf\hspace*{-12 pt}
\mbox{\hspace*{-6pt}Measurement of the Branching Fractions} \\
and Forward-Backward Asymmetries
\vspace*{5mm}
of the Z{\boldmath $^0$} into Light Quarks \\
}
\end{center}
\bigskip
\begin{center}
{\LARGE The OPAL Collaboration}
\end{center}
\bigskip
\begin{center} {\large\bf Abstract} \end{center}
\noindent 
Using approximately 4.3 million hadronic $\rm{Z}^0$ decays collected with the 
OPAL detector at LEP between 1990 and 1995, we measure the branching fractions 
of the $\rm{Z}^0$ into up-type and down-type light quarks, $R_q$, and the 
forward-backward asymmetries, $A_{FB}(q)$, using high-momentum stable particles 
as a tag.  Adopting a method that employs double tagged events to determine the 
flavour tagging efficiencies, and assuming the flavour independence of strong
interactions and SU(2) isospin symmetry, we measure:
\begin{eqnarray*}
\frac{R_{\rm d,s}}{(R_{\rm{d}}+R_{\rm{u}}+R_{\rm{s}})} & \ = \ & 
0.371 \ \pm 0.016 \ \rm{(stat.)} \  \pm 0.016 \ \rm{(syst.)} \ 
\rm{and} \\
A_{FB}{\rm (d,s)} & \ = \ &
0.068 \ \pm 0.035 \ \rm{(stat.)} \  \pm 0.011 \ \rm{(syst.)} \ , 
\end{eqnarray*}
assuming the branching fractions and forward-backward asymmetries of down and 
strange quarks to be equal.  The results are essentially free of assumptions
based on hadronisation models.  These results are in agreement with the 
Standard Model expectations and are used to infer the left and right handed 
couplings of strange/down quarks to the $\rm Z^0$, yielding
$$
g_L^{\rm d,s} \ = \ -0.44^{+0.13}_{-0.09} \ \ \ {\rm and} \ \ \
g_R^{\rm d,s} \ = \ +0.13^{+0.15}_{-0.17} \ .
$$
The results for the up quark, $R_{\rm{u}}/(R_{\rm{d}}+R_{\rm{u}}+R_{\rm{s}})=   
0.258\pm 0.031 \,\rm{(stat.)} \pm 0.032\,\rm{(syst.)}$ and
$A_{FB}({\rm{u}})= 0.040 \pm 0.067\,\rm{(stat.)} \pm 0.028\,\rm{(syst.)}$,
are fully negatively correlated and almost completely positively correlated, 
respectively, with the corresponding down-type results.
\bigskip
\bigskip
\bigskip
\begin{center}{\large
(Submitted to Zeitschrift f\"ur Physik C)
}\end{center}
\bigskip
\bigskip
\end{titlepage}
%%%%%%%%%%%%%%%%%%%%%%%%%%%%%%%%%%%%%%%%%%%%%%%%%%%%%%%%%%%%%%%%%%%%%%%%%%%%%%%%
\begin{center}{\Large        The OPAL Collaboration
}\end{center}\bigskip
\begin{center}{
%begin authorlist
K.\thinspace Ackerstaff$^{  8}$,
G.\thinspace Alexander$^{ 23}$,
J.\thinspace Allison$^{ 16}$,
N.\thinspace Altekamp$^{  5}$,
K.J.\thinspace Anderson$^{  9}$,
S.\thinspace Anderson$^{ 12}$,
S.\thinspace Arcelli$^{  2}$,
S.\thinspace Asai$^{ 24}$,
D.\thinspace Axen$^{ 29}$,
G.\thinspace Azuelos$^{ 18,  a}$
% ...
%end authorlist
}\end{center}\bigskip
\bigskip
%begin institutes
% ...
$^{  2}$Dipartimento di Fisica dell' Universit\`a di Bologna and INFN,
I-40126 Bologna, Italy
\newline
% ...
$^{  5}$Cavendish Laboratory, Cambridge CB3 0HE, UK
\newline
% ...
$^{  8}$CERN, European Organisation for Particle Physics,
CH-1211 Geneva 23, Switzerland
\newline
$^{  9}$Enrico Fermi Institute and Department of Physics,
University of Chicago, Chicago IL 60637, USA
\newline
% ...
$^{ 12}$Indiana University, Department of Physics,
Swain Hall West 117, Bloomington IN 47405, USA
\newline
% ...
$^{ 16}$Department of Physics, Schuster Laboratory, The University,
Manchester M13 9PL, UK
\newline
% ...
$^{ 18}$Laboratoire de Physique Nucl\'eaire, Universit\'e de Montr\'eal,
Montr\'eal, Quebec H3C 3J7, Canada
\newline
% ...
$^{ 23}$Department of Physics and Astronomy, Tel Aviv University,
Tel Aviv 69978, Israel
\newline
$^{ 24}$International Centre for Elementary Particle Physics and
Department of Physics, University of Tokyo, Tokyo 113, and
Kobe University, Kobe 657, Japan
\newline
% ...
$^{ 29}$University of British Columbia, Department of Physics,
Vancouver BC V6T 1Z1, Canada
\newline
% ...
$^{  a}$ and at TRIUMF, Vancouver, Canada V6T 2A3
\newline
% ...
$^{  d}$ and Department of Experimental Physics, Lajos Kossuth
University, Debrecen, Hungary
\newline
% ...
\newline
%end notes

%%%%%%%%%%%%%%%%%%%%%%%%%%%%%%%%%%%%%%%%%%%%%%%%%%%%%%%%%%%%%%%%%%%%%%%%%%%%%%%%
\newpage
\section{Introduction}
 
One of the basic assumptions of the Standard Model~\cite{bib-SM} is flavour 
universality, namely that apart from mass effects, the gauge couplings of all
fermions depend only on their charge, weak isospin, and colour.  While this 
assumption has been experimentally tested to high accuracy in the charged 
lepton sector, results on quarks are generally less precise.  The large number 
of $\rm{Z}^0$ decays recorded at the LEP $\rm e^+e^-$ collider opens 
the possibility for high-precision measurements of the $\rm{Z}^0$ 
couplings to several individual fermion species.  Accurate measurements exist 
particularly for the bottom quarks, and to a lesser extent for the charm 
quark~\cite{bib-PDG}. However, at LEP, few measurements exist of the couplings of 
individual up, down, and strange quarks to the neutral weak current.  The 
couplings of up and down quarks have been obtained at low $Q^2$ from 
lepton-nucleon scattering and from atomic parity violation and are in agreement
with the Standard Model expectations~\cite{bib-leptnucl}.  Direct measurements 
of light flavours on the $\rm{Z}^0$ resonance are scarce.  Apart from the 
overall hadronic width, the yield of photon radiation from quarks is 
particularly sensitive to up-type quarks~\cite{bib-qqgamma}.  A first 
measurement of the strange quark forward-backward asymmetry, which is 
dependent on hadronisation models, has been published in~\cite{bib-DELFISTR}.
 
In this paper we determine the decay branching fractions of the $\rm{Z}^0$ 
into up-type and down-type light quarks and the forward-backward asymmetries 
with only few model assumptions, as introduced in~\cite{bib-letts+mattig}.  The
method relies on the property that a high-energy particle in a jet has a 
flavour correlation with the primary quark.  Since charm and bottom quarks do 
not contribute much to the production of long-lived hadrons with a significant 
fraction of the beam energy, a selection of events with $\pi^\pm$, $\rm K^\pm$, 
$\rm p(\overline{{p}})$, $\rm K^0_S$, or $\Lambda(\overline{\Lambda})$ baryons 
of high energy provides rather pure samples of light flavours\footnote{Unless 
explicitly stated otherwise, charge conjugation of the tagging particle and 
primary quark types is implied throughout this paper.}.  The main challenge is 
to determine the relative yields of up, down, and strange quarks in such event 
samples.  This is achieved by using the information from double tagged events, 
where two high-momentum particles are found in opposite event hemispheres, 
together with some general hadronisation symmetries.
 
The method of measuring the branching fractions and the forward-backward 
asymmetries is described in detail in Section~2.  The elements of the 
OPAL detector pertinent to this analysis and the basic event selection are
presented in Sections~3 and~4.  In Section~5 we describe the estimation of the 
charm and bottom backgrounds in our high-momentum samples and in Section~6 the 
electroweak observables are determined for up-type and down-type light quarks.
Systematic uncertainties are evaluated in Section~7 and the conclusions are 
given in Section~8.
 
\section{The Method}
 
The observed long range charge correlations in $\rm e^+e^-$ 
events~\cite{bib-QCORREL} show that the particle with the highest energy in a 
jet tends to carry the quantum numbers of the primary quark.  Thus, jets from 
primary strange quarks lead to high-energy strange hadrons, for example, and 
primary up and down quarks lead to high-energy pions and protons.  We base our
analysis on this property and tag light quark events by identifying high-energy 
charged pions, charged and neutral kaons, protons, and $\Lambda$ baryons with 
$x_p=2p_h/E_{\rm cm}\ge 0.5$, where $p_h$ is the momentum of the tagging 
hadron, $h$, and $E_{\rm cm}$ is the centre-of-mass energy of the event.  The 
value of $x_p>0.5$ is chosen to minimise the statistical and systematic 
uncertainties.  We now make a general discussion of the method without 
considering the detector.

\subsection{Method to Determine the Branching Fractions}
 
If the relations between particle types and primary flavours were unambiguous, 
double tagged events, in which a tagging hadron is found in each of the event 
hemispheres, could be used to determine the flavour tagging efficiencies in a
straightforward way.  The method would then be essentially free from 
uncertainties due to the detailed properties of the flavour tag and detector 
effects.  Almost no ambiguity exists for bottom particles, where the double 
tagging method has been successfully applied (see~\cite{bib-Rb}, for example).  
On the other hand, high-energy light-flavour mesons and baryons can be produced 
by processes other than the hadronisation of primary quarks.  Leading mesons 
contain a quark and an antiquark, either of which could be the quark into which 
the $\rm{Z}^0$ decayed directly, which introduces some ambiguity. In addition, 
decays of these so-called primary hadrons and of particles produced from the 
hadronisation sea tend to obscure further the primary flavour source of each 
particle type.  Thus, each particle species is produced from a mixture of 
several primary quark flavours.  Neglecting correlations, the number of tagged 
event hemispheres (as defined by the thrust axis) and the number of double 
tagged events can be expressed as:
\begin{eqnarray}
{N_h \over N_{\rm had} } & \ = \ &
2 \hspace*{-2mm} \sum_{q={\rm d,u,s,c,b}} \eta_q^h  R_q 
{\ \ \rm{and}} \\
{N_{h h'} \over N_{\rm had} } & \ = \  &
\sum_{q={\rm d,u,s,c,b}} \eta_q^h  \eta_q^{h'} R_q \ , 
\end{eqnarray}
where $N_h$ is the number of hemispheres with a tagging hadron $h=\pi^\pm,
\rm K^\pm,p(\overline{\rm p}),\rm K^0_S, \Lambda(\overline{\Lambda})$, 
$N_{\rm had}$ is the number of hadronic $\rm{Z}^0$ decays, and 
$N_{h h'}$ is the number of double tagged events with tagging 
particle types $h$ and $h'$. The $\eta_q^h$ denote the fraction of hemispheres 
with a primary quark flavour $q$ which are tagged by a hadron of type $h$, and 
$R_q$ is the hadronic branching fraction of the $\rm{Z}^0$ to quarks $q$:
\begin{equation}
R_q \ = \ \frac{\Gamma_{{\rm Z}^0\rightarrow q\bar q}}{\Gamma_{\rm had}}.
\end{equation}

As will be shown in Section~5, the charm and bottom fractions can be 
determined separately in a straightforward way, leaving fifteen unknown 
$\eta^h_q$ and three unknown $R_q$.  On the other hand, for the five tagging
particle types we consider, there is a system of five equations for tagged
hemispheres (Eq.~1) and fifteen equations for double tagged event types 
(Eq.~2).  Due to the non-linearity and degeneracy of the equation system, 
it is not solvable and additional constraints have to be found to obtain a 
solution.

As discussed in detail in~\cite{bib-letts+mattig}, the necessary constraints 
can be derived from  SU(2) isospin symmetry and the flavour independence of 
QCD.  We use the following hadronisation relations:
\begin{eqnarray}
\eta_{\rm{d}}^{\pi^\pm}     \ & = & \ \eta_{\rm{u}}^{\pi^\pm} \ ,\\
\eta_{\rm{d}}^{\rm K^0(\overline{K}^0)} \ & = & \ 
\eta_{\rm{u}}^{\rm K^\pm} \ ,  \\
\eta_{\rm{u}}^{\rm K^0(\overline{K}^0)} \ & = & \ 
\eta_{\rm{d}}^{\rm K^\pm} \ ,  \\
\eta_{\rm{s}}^{\rm K^0(\overline{K}^0)} \ & = & \ 
\eta_{\rm{s}}^{\rm K^\pm} \ , \ \rm{and} \\
\eta_{\rm{u}}^{\Lambda(\overline{\Lambda})} \ & = & \ 
\eta_{\rm{d}}^{\Lambda(\overline{\Lambda})} \ .
\end{eqnarray}
Note that $\rm K^0(\overline{K}^0)$ implies $\rm K^0_S$ plus $\rm K^0_L$.  
Small deviations from these relations are discussed later in Section~7.  In 
addition, the overall normalisation $\sum R_q=1$ provides another constraint:
\begin{equation}
R_{\rm{u}} \ + \ R_{\rm{d}} \ + \ R_{\rm{s}} \ 
= \ 1 \ - \ R_{\rm{c}} \ - \ R_{\rm{b}} \  \ = \ 0.620 \pm 0.010 \ , 
\end{equation}
where the LEP measurements~\cite{bib-PDG} of $R_{\rm{b}}$ and $R_{\rm{c}}$ 
can be used to constrain the sum of the light-flavour branching 
fractions.  To be independent of the measurements of the heavy flavour 
fractions, we also express our results in terms of:
\begin{equation}
R_q' \ = \ { \Gamma_{q\bar{q}} \over
\Gamma_{\rm d\bar{d}}+\Gamma_{\rm u\bar{u}}+ \Gamma_{\rm s\bar{s}}} 
\ = \ { R_q \over R_{\rm{d}}+R_{\rm{u}}+R_{\rm{s}} }
\end{equation}
The $R'_q$ are related to the $R_q$ via $R_q=R'_q(1-R_{\rm b}-R_{\rm c})$.

These constraints are still not sufficient to allow the system to be solved.  
Motivated by the weak isospin structure of the Standard Model, we assume 
$R_{\rm{d}}=R_{\rm{s}}$, thereby reducing the number of unknown $R_q$ 
and making possible a solution of the equation system.  
 
\subsection{Method to Determine the Forward-Backward Asymmetries}

The other directly measurable electroweak-related observable is the 
forward-backward asymmetry $A^h_{FB}$ of a hadron $h$:
\begin{equation}
A_{FB}^h \ = \ {N_h(\cos\theta >0) \ - \ N_h(\cos\theta<0) \over N_h} \ ,
\end{equation}
where $\theta $ is the angle of the tagging hadron $h$ with respect to the 
incoming electron direction.  The relation between this directly observable 
asymmetry and the desired forward-backward asymmetries of the quarks is 
given by~\cite{bib-letts+mattig}:
\begin{equation}
A_{FB}^{h}  \ = \ \sum_q \left\{ s_q  f_q^{h}
 (2r^{h}_q - 1) \right\}  A_{FB}(q)  \ ,
\end{equation}
where the reliabilities of the charge tag are given by 
$r^h_q=N_{h,q}^{\rm correct}/N_{h,q}$, $N_{h,q}^{\rm correct}$ 
being the number of 
hadrons $h$ which have the same sign of the charge as the primary quark $q$.
The reliability therefore takes into account dilutions due to the 
misidentification of the sign of the charges of the quarks, $s_q$.  Finally, 
$f_q^h$ is the fraction of tagged hadrons $h$ stemming from a primary quark 
$q$, i.e. $f_q^h=(\eta_q^h R_q)/(\sum_{q'}\eta^h_{q'}R_{q'})$.  Therefore, the 
forward-backward asymmetries $A_{FB}(q)$ can be determined only after the 
$\eta_q^h$, $R_q$, and reliabilities $r_q^h$ are known.  The reliabilities can 
be determined from the ratio of double tagged events with tagging particles of 
opposite charge, $N_{h h'}^{\rm OPP}$ over the total number of double 
tagged events, $N_{h h'}$:
\begin{equation}
{ N_{h h'}^{\rm OPP} \over N_{h h'} } \  = \  
\sum_q    {\eta_q^h  \eta_q^{h'}  R_q \over
(\sum_{q'} \eta_{q'}^h \eta_{q'}^{h'} R_{q'}) }
\ \Bigl\{ r^h_q  r^{h'}_q \ + \ (1-r^h_q) (1-r^{h'}_q) \Bigr\} \ .
\end{equation}

Since the $\rm K^0_S$ provides no information on the charge of the primary 
quark, we restrict the determination of the forward-backward asymmetries to 
charged pions, charged kaons, protons, and $\Lambda$ baryons.  With these four 
hadron types we obtain four measurements of $A_{FB}^h$ and ten ratios 
$N_{h h'}^{\rm OPP}/N_{h h'}$ which must be used to 
determine 12 unknown light flavour reliabilities and the three asymmetries 
$A_{FB}(q)$.  Here we assume that the heavy flavour terms can be determined
separately, as will be shown later in Section~5.  As for the 
solution of the equation system for the branching fractions, we invoke 
hadronisation symmetries based on SU(2) isospin invariance to reduce the number 
of unknown reliabilities:
\begin{eqnarray}
r_{\rm{d}}^{\pi^\pm} \ & = & \ r_{\rm{u}}^{\pi^\pm} \ , \\
r_{\rm{s}}^{\pi^\pm} \ & = & \ 0.5 \ , \ {\rm and} \\
r_{\rm{d}}^{\Lambda(\overline{\Lambda})}   \ & = & \ 
1-r_{\rm{u}}^{\Lambda(\overline{\Lambda})}  \ . 
\end{eqnarray}
Note that for the baryons (the proton and $\Lambda$), the tagging hadron
carries the sign of the baryon number of the primary quark, not the 
electric charge, so that $r_q^{\rm baryon} < 0.5$ for down-type quarks
and $r_q^{\rm baryon} > 0.5$ for up-type quarks. In addition, in order to
solve the equation system we must assume that (see~\cite{bib-letts+mattig}):
\begin{equation}
r_{\rm{d}}^{\rm K^\pm} \ = \ 0.20 \pm 0.10 \ ,
\end{equation}
where the value is taken from the JETSET model~\cite{bib-JETSET} 
and is assigned a large error to take into account uncertainties in the 
JETSET modelling.
This particular reliability is chosen as the one to be fixed since 
$f_{\rm{d}}^{\rm K^\pm}$ is small compared to the other flavour fractions.  
Therefore, even a large uncertainty in $r_{\rm{d}}^{\rm K^\pm} $ does not 
affect significantly the final result.
 
In the preceding discussion we have ignored biases due to geometrical and 
kinematic constraints which have to be taken into account.  Requiring a 
high-energy hadron in an event reduces the phase space for gluon 
bremsstrahlung and thus introduces a kinematic correlation between the 
event hemispheres.  Similarly, restrictions on the geometrical acceptance 
introduce corrections to the equation systems we use.  We parametrise this 
correction by a factor $\rho$, so that the double tagging probability,
$\eta_q^{h h'}$, is given by:
\begin{equation}
\eta_q^{h h'} \ = \ \rho  \eta_q^h \eta_q^{h'} \ .
\end{equation}
In the ideal case of no bias, $\rho=1$.  Model calculations 
suggest~\cite{bib-letts+mattig} that this correlation is independent of the 
tagging particle type and the primary quark flavour and is approximately 1.07 
for a sample of tagging hadrons with $x_p>0.5$ without experimental cuts and 
neglecting detector effects, and is essentially due to gluon bremsstrahlung.
In order for the measurement of the light-flavour electroweak parameters to be 
as model independent as possible, we let $\rho$ be a free parameter in the 
equation system but assume that it is independent of the tagging hadron species.
Note that $\rho$ is largely uncorrelated with the other parameters. 
 
\section{The OPAL Detector}
 
The OPAL detector is described in detail in~\cite{bib-OPALDET}.  Here we 
summarise only the features of the detector which are important for this 
analysis.
 
Central to this study is the determination of the momentum and the identity of 
different types of charged hadrons and neutral hadrons which decay to stable 
charged particles.  OPAL has a system of tracking devices inside a solenoid
which provides a magnetic field of 0.435{\,}T.   A charged track momentum 
resolution of $\sigma_p/p=0.02 \oplus 0.0015~p_t$, where $p_t$ is the component 
of the total track momentum $p$ in the plane perpendicular to the beam axis 
measured in GeV, has been achieved.  The innermost part is a silicon 
microvertex detector~\cite{bib-OPALSI}, surrounded by three drift chambers: a 
vertex detector, a large volume jet chamber which provides up to 159 space 
points per track, and $z$-chambers to give a more precise measurement of the 
polar angle of charged tracks\footnote{ OPAL uses a right handed coordinate 
system in which the $z$ axis points along the direction of the electron beam, 
$r$ is the coordinate normal to this axis, and $\theta$ and $\phi$ are the 
polar and azimuthal angles with respect to $z$.}.  The large 
number of measurements in the jet chamber also provides a determination of 
the specific ionisation energy loss, d$E$/d$x$, with a resolution of 
$\sigma({\rm{d}}E /{\rm{d}}x)/({\rm{d}}E/{\rm{d}}x)\sim 
0.035$~\cite{bib-DEDXOPAL} for well-separated tracks with $|\cos\theta|<0.7$.  
This resolution allows the identification of charged pions, charged kaons, and 
protons up to the highest particle momenta~\cite{bib-OPALHADR}.  The large 
radius of the jet chamber ($R=185$~cm) also allows a high reconstruction 
efficiency for large-momentum, weakly decaying hadrons with relatively long 
decay lengths, such as ${\rm K^0_S}\to\pi^+\pi^-$ and 
$\Lambda\to {\rm p}\pi^-$.
 
To estimate the contributions from bottom quarks, we also identify secondary 
vertices from b hadron decays~\cite{bib-VERTXTAG} and use electron and muon 
identification~\cite{bib-eid,bib-muid} to tag semi-leptonic bottom decays.
The identification of a secondary vertex profits particularly from the
high-precision measurements in the silicon microvertex detector.  The lepton 
identification is largely based on the electromagnetic calorimeter, which 
consists of 11\,704 lead glass blocks which subtend a solid angle of
$40\times40$~mrad, and muon chambers which are placed behind an average of 
eight absorption lengths of detector material. 
 
To determine detector efficiencies and possible detector biases, 
we use a sample of approximately eight million simulated hadronic Z$^0$ 
decays generated with the JETSET model~\cite{bib-JETSET} and passed through 
a detailed simulation of the OPAL detector~\cite{bib-GOPAL}.  The fragmentation 
parameters have been tuned to describe overall event shapes and distributions
as described in detail in~\cite{bib-toon}.
Events are generated with two versions of JETSET. Events with JETSET~7.3 use
a simulation of the detector up to and including 1993, while the 
JETSET~7.4 events use a simulation of the detector corresponding to
subsequent years.
 
\section{Event Selection}

The analysis is based on approximately 4.3 million multihadronic Z$^0$ decays 
collected between 1990 and 1995.  Of these events approximately 90\% were 
collected at centre-of-mass energies within $\pm 200$~MeV of the $\rm{Z}^0$ 
mass, and the rest of the events within $\pm 3$~GeV above and below 
the Z$^0$ peak.  The standard OPAL requirements for the multihadronic event
selection are given in~\cite{bib-MHSELECT}.  To enrich the light-flavour 
fraction we retain only those events which have at least one well measured 
charged track, a ${\rm K^0_S}$, or a $\Lambda$ baryon, with a scaled momentum 
$x_p>0.5$.  Details of the particle selections are given below.  Since this 
high-$x_p$ cut rejects relatively few $\rm{Z}^0\to\tau^+\tau^-$ events in 
our sample, we further demand at least eight good charged 
tracks~\cite{bib-goodtracks} in an event.  In addition, to ensure good 
$\pi^\pm$, $\rm K^\pm$ and proton separation, and reliable 
$\rm K^0_S$ and $\Lambda$ reconstruction, we require that the tagging 
particles have a 
polar angle $|\cos\theta|<0.7$.  For $|\cos\theta|>0.7$, the d$E$/d$x$ 
separation of kaons and protons from pions degrades rapidly with increasing 
$|\cos\theta|$ of the tracks.  In order to select events which are well 
contained in the detector, we restrict the polar angle of the thrust axis 
(calculated using both charged tracks and clusters in the electromagnetic
calorimeter which have no associated track in the jet chamber) to satisfy 
$|\cos\theta_{\rm Thrust}|<0.8$.  With these requirements, 198\,309 events are 
retained. The remaining background from $\rm{Z}^0\to\tau^+\tau^-$ events in 
our sample is negligible (less than 0.05\%), as estimated from Monte Carlo 
simulations.  

Crucial to this analysis is the identification and separation of samples of
charged pions, charged kaons and protons, which is achieved using the 
d$E$/d$x$ measurement, and ${\rm K^0_S}$ and $\Lambda$ baryons, which are 
observed by reconstructing their decay vertices and calculating the invariant 
mass of the decay products.  Efficiency losses which are common to all particle 
types, such as the finite geometric acceptance, etc., can be absorbed into the 
values of $\eta_q^h$.  However, other sources, such as those due to particle 
identification (d$E$/d$x$, secondary vertex finding) have to be taken into 
account for the hadronisation symmetries.  Similarly, there exists some 
misidentification probability leading to a migration from the true to an 
apparent particle identity, which is especially relevant for the stable charged 
particles.  This must also be taken into account and the hadronisation 
symmetries corrected for different detection efficiencies and migrations.  Due 
to the fundamentally different reconstruction of the weakly decaying 
$\rm K^0_S$ and $\Lambda$, these samples are largely decoupled from the 
charged hadron samples.  The relation between apparent particle type 
$h_{\rm det}$ and true particle type $h$ is given by some flow matrix, 
${\cal E}^{h}_{h_{\rm det}}$, such that
\begin{equation}
N_{h_{\rm det}} \ = \ \sum_h {\cal E}^{h}_{h_{\rm det}} N_h \ .
\end{equation}
Similarly, the forward-backward asymmetries for the measured samples 
$A^{h_{\rm det}}_{FB}$ are related to the pure particle type asymmetries 
$A^{h}_{FB}$ by
\begin{equation}
A^{h_{\rm det}}_{FB} \ = \ \sum_h {\cal E}^{h}_{h_{\rm det}} A^{h}_{FB} \ .
\end{equation}
Apart from reducing the discrimination power between the various flavours,
some systematic uncertainties related both to the efficiency and purity
are introduced.
 
For the d$E$/d$x$ measurement, tracks are required to have at least 20 hits in 
the jet chamber used in the calculation of the energy loss, a polar angle 
$\theta$ satisfying $|\cos\theta |<0.7$, a distance of closest 
approach to the interaction point in the plane orthogonal to the beam direction 
of $|d_0|<2$~mm, and the corresponding distance along the beam direction 
$|z_0|<40$~cm.  To suppress badly measured tracks we further restrict the 
scaled momentum to $x_p<1.07$, which takes into account the 7\% momentum 
resolution of a track with $x_p=1$.  A study of muon pair events finds that 
less than 0.3\% of tracks which have the full beam energy have an incorrect 
charge assignment.

To separate samples of pions, kaons and protons, we use for each track the 
d$E$/d$x$ weight, $w_h$, which is defined as the signed $\chi^2$ probability of 
the track to be consistent with a certain particle species hypothesis, $h$.  
The sign represents the sign of the difference between the measured 
d$E$/d$x$(meas.) and the expected d$E$/d$x$(exp.) for the particle species 
hypothesis, namely a positive weight if d$E$/d$x$(meas.)$>$d$E$/d$x$(exp.) and 
a negative weight if d$E$/d$x$(meas.)$<$d$E$/d$x$(exp.).  In particular, we 
require:
\begin{itemize}
\item for pion candidates:   
               ($w_{\pi^{\pm }} > 0.01$ or  $w_{\pi ^{\pm }}<-0.1$), and 
                    $|w_{\rm K^{\pm}}|<0.1$ \ ;
\item for kaon candidates:   
               $|w_{\rm K^{\pm}}|>0.1$ and $|w_{\pi ^{\pm}}|<0.1$ \ ; 
\item for proton candidates: 
              ($w_{\rm p( \overline{\rm p})}>0.1$ or 
               $w_{\rm p( \overline{\rm p})}<-0.01$), 
                    and $|w_{\rm K^{\pm }}|<0.1$ \ .
\end{itemize}
These selection criteria result in three disjoint track samples.
The d$E$/d$x$ separation power as a function of charged track momentum
is shown in Fig.~1.
In addition, charged tracks are rejected if they pass either the standard 
electron or muon identification requirements~\cite{bib-eid,bib-muid}.  
After these requirements, backgrounds from electrons and muons
are negligible, as determined from Monte Carlo simulations.
The largest contamination is the 0.8\% of pion candidates which 
are actually muons.  In addition, some contamination from charged hyperons 
(mostly $\Sigma^-$) is present in the proton sample at the 10\% level 
according to the Monte Carlo simulations.  However, since we make no 
hadronisation assumptions about the proton sample, this contamination is 
not a problem for the analysis.   

These cuts (including the event and thrust axis cuts) lead to efficiencies of 
$(34.7 \pm 0.1 \pm 5.2)\%$ for pion tagged hemispheres, 
$(29.5 \pm 0.1 \pm 4.6)\%$ for kaon tagged hemispheres and
$(23.4 \pm 0.3 \pm 5.3)\%$ for proton tagged hemispheres, where in each case 
the uncertainties are statistical and systematic.  The efficiency
is defined as the number of event hemispheres which are tagged at the 
detector level divided by the number of event hemispheres which are tagged
at the Monte Carlo generator level.  Differences in the tagging efficiencies 
are due entirely to the d$E$/d$x$ selections.  The systematic 
uncertainty is determined by scaling the corresponding widths of the d$E$/d$x$ 
distributions, $\sigma$, by $1.00 \pm 0.05$ and by varying the expected 
d$E$/d$x$ value for each particle type by $\pm0.15\,\sigma$.  The ranges of 
these uncertainties come from checks with pions from ${\rm K^0_S}\to\pi^+\pi^-$ 
decays and protons from $\Lambda\to{\rm{p}}\pi^-$ decays for track momenta 
greater than 2~GeV in which the means and widths of the normalised d$E$/d$x$ 
distributions showed maximum deviations of $\pm0.15\,\sigma$ and $\pm 5\%$, 
respectively.  The uncertainties are therefore strongly positively correlated 
between the various particle species.  The other selection criteria were 
studied in the Monte Carlo event samples and found to have no significant 
bias to select any one tagging particle type over another.

The procedures to identify the weakly decaying hadrons $\rm K^0_S$ and 
$\Lambda$ are described in detail in~\cite{bib-OPALK0S} 
and~\cite{bib_OPALLAMBDA}, respectively.  Here we summarise the main 
ingredients.  To find ${\rm{K^0_S}}\to\pi^+\pi^-$ and $\Lambda\to 
{\rm{p}}\pi^-$ candidates, we combine two oppositely charged tracks which have 
at least 20 hits in the jet chamber.  We then search for a crossing point of 
these tracks in the plane orthogonal to the beam axis.  If a good secondary 
vertex is found, the $\pi^+\pi^-$ and $ {\rm{p}}\pi^-$ invariant masses of the 
combinations are calculated.  Good $\rm K^0_S$ candidates are required to have 
$x_p>0.5$ and have invariant masses in the ranges 430~MeV~$<m_{\pi^+\pi^-}
<$~570~MeV and $m_{ {\rm{p}}\pi^-} > 1.13$~GeV, in order to reduce the 
contamination from $\Lambda\to{\rm{p}}\pi^-$ decays, where $m_{\pi^+\pi^-}$ is 
the invariant mass of the tracks assuming that they are a pair of pions and 
$m_{ {\rm{p}}\pi^-}$ a proton-pion pair.  Likewise, all candidates which have 
$x_p>0.5$ and 1.10~GeV~$<m_{ {\rm{p}}\pi^-}<$~1.13~GeV are accepted as 
high-momentum $\Lambda$ candidates.  No additional rejection of 
${\rm{K^0_S}}\to\pi^+\pi^-$ background in the $\Lambda$ sample using an 
invariant mass cut is made since such a rejection cannot be done efficiently.  
The invariant mass distributions for all candidates which pass these selection 
criteria are shown in Figs.~2 and~3.  The detection efficiencies are determined 
using the Monte Carlo simulated events.  After correcting for mass resolution 
differences between data and Monte Carlo simulation as 
in~\cite{bib_OPALLAMBDA}, the efficiencies, defined as before, are found to 
be $(9.4 \pm 0.1 \pm 0.3)\%$ for $\rm K^0_S$ tagged hemispheres and 
$(4.7 \pm 0.1 \pm 0.3)\%$ for $\Lambda$ tagged hemispheres, 
where the errors are statistical and systematic, 
respectively.  The systematic errors are mainly due to uncertainties in the 
correction factors for the different mass resolutions in the data and Monte 
Carlo and the simulation of other cut distributions.  Detailed treatments of 
the determination of these systematic errors can be found 
in~\cite{bib-OPALK0S,bib_OPALLAMBDA}.  With these requirements we select the 
number of tagged hemispheres and double tagged events listed in Table~1.

The purities of the final samples, as determined from the efficiencies 
calculated above with appropriate weighting by the relative particle yields 
in the data are given in Table~2.  Although there is some uncertainty in the 
particle identification efficiencies, the charged pion, kaon and proton 
sample purities are known to $\pm 0.011$, $\pm 0.013$ and $\pm 0.052$, 
respectively.  The errors are smaller for the purities than for the 
efficiencies because of the positive correlations among the efficiencies.  
The $\rm K^0_S$ and $\Lambda$ sample purities are largely uncorrelated with 
those of the charged hadron samples.  In addition to small combinatorial 
backgrounds composed of randomly paired tracks, ${\rm{K^0_S}}\to\pi^+\pi^-$ 
decays form the principal background in the $\Lambda$ sample, and $\Lambda\to 
\rm{p} \pi^-$ combinations form a somewhat less important background in 
the $\rm K^0_S$ sample.  The total backgrounds are determined by fitting to 
the invariant mass spectrum shown in Fig.~2, as in~\cite{bib-OPALK0S}, and by
using a sideband method to determine the background in the distribution shown 
in Fig.~3, as in~\cite{bib_OPALLAMBDA}.  For the $\rm K^0_S$ sample, after 
subtracting the contribution from $\Lambda\to  {\rm{p}}\pi^-$ decays, a 
combinatorial background level of $(5.6 \pm 2.9)\%$ is found, where the error 
is systematic~\cite{bib-OPALK0S}.   Combining this with the relative 
uncertainty on the efficiency, the total relative systematic error on the 
$\rm K^0_S$ yield is $\pm 4.1\%$.  For the $\Lambda$, the combinatorial 
background present in the sample is $(7.7\pm 2.4)\%$, where the error is again 
systematic~\cite{bib_OPALLAMBDA}.  The total relative error on the $\Lambda$ 
yield is therefore $\pm 6.3\%$.
 
\begin{table}[tbh]
\renewcommand{\arraystretch}{1.2}
\begin{center}
\begin{tabular}{|c||r|r|r|r|r|r|} \hline
hadron    & \multicolumn{1}{c|}{tagged} 
& \multicolumn{5}{c|}{double tagged events} \\
type      
& hemispheres   
& $\pi^\pm$  
& $\rm K^\pm$ 
& $\rm p(\overline{p})$ 
& $\rm K^0_S$ 
& $\Lambda(\overline{\Lambda})$ \\ 
\hline\hline
$\pi^\pm$                     & 52\,170 & 392 & 416 &  99 & 46 & 18   \\ \hline
$\rm K^\pm$                   & 40\,229 &     & 265 & 136 & 48 & 15   \\ \hline
$\rm p(\overline{p})$         &  9\,350 &     &     &  13 & 15 &  4   \\ \hline
$\rm K^0_S$                   &  5\,026 &     &     &     &  3 &  1   \\ \hline
$\Lambda(\overline{\Lambda})$ &  1\,349 &     &     &     &    &  1   \\ \hline
\end{tabular}
\caption{Number of tagged event hemispheres and double tagged events 
for $x_p>0.5$.}
\end{center}
\renewcommand{\arraystretch}{1.0}
\end{table}
 
\begin{table}[tbh]
\renewcommand{\arraystretch}{1.2}
\begin{center}
\begin{tabular}{|c||c|c|c|c|c|} \hline
assigned & true $\pi^\pm$ & true $\rm K^\pm$  
& true $\rm p(\overline{p})$   
& true $ \rm K^0_S$ 
& true $\Lambda(\overline\Lambda)$ \\ \hline\hline
$\pi^\pm$             & 0.895 & 0.080 & 0.002 & 0.011 & 0.000 \\ \hline
$\rm K^\pm$           & 0.106 & 0.712 & 0.171 & 0.001 & 0.008 \\ \hline
$\rm p(\overline{p})$ & 0.013 & 0.367 & 0.591 & 0.001 & 0.028 \\ \hline
$\rm K^0_S$           & 0.003 & 0.001 & 0.000 & 0.979 & 0.017 \\ \hline
$\Lambda(\overline\Lambda)$  
                      & 0.001 & 0.003 & 0.000 & 0.307 & 0.690 \\ \hline
\end{tabular}
\caption{The particle composition in samples assigned as charged pions,
charged and neutral kaons, protons, $\Lambda$ baryons.  Note that 
in some of the rows the numbers do not add up to 100\%, due to the presence 
of small backgrounds from other sources such as leptons.  The combinatorial 
backgrounds under the $\rm K^0_S$ and $\Lambda$ signals are not considered 
here and are discussed in the text.  Charged hyperons were treated as 
``protons'' for the purposes of identification.}
\end{center}
\renewcommand{\arraystretch}{1.0}
\end{table}

\section{Heavy Quark Contributions}

Before determining the light-flavour electroweak properties, we first find the 
contributions of charm and bottom quarks to the selected samples.  In the case 
of the bottom quark, we determine the fraction directly from the data 
(except for the tagged $\Lambda$ sample) by 
searching for a secondary vertex or for a lepton with a high transverse 
momentum relative to the jet direction in the high-$x_p$ samples.  Lack of 
statistics prevents a similar method from being used to determine the charm 
backgrounds with the necessary precision. Instead we use the JETSET model and 
estimate the uncertainties based on the measured production and decay 
properties of charmed hadrons.  Due to the low statistics available, the b 
quark flavour fractions and reliabilities for the tagged $\Lambda$ sample
are also taken from the simulation.

\subsection{Bottom Quark Contributions}
 
To determine the fractions $f_{\rm{b}}^h$ of $\rm Z^0\to b\overline{b}$ 
events in the high-$x_p$ event samples, we search for a secondary vertex 
displaced from the primary event vertex in the event hemisphere opposite to 
that of the high-$x_p$ tagging particle.  The details 
of the vertex finding are discussed in~\cite{bib-VERTXTAG}.  We require the 
vertex to have at least four assigned tracks and to have a decay length 
significance $l/\sigma_l>$10, where $l$ is the observed distance to the event 
vertex and $\sigma_l$ its error.  This yields a b tagging efficiency per event 
hemisphere, $\epsilon_{\rm{b}}$, of about 16\% with a purity, 
$P_{\rm{b}}$, of $(96\pm 1)\%$, using the methods in~\cite{bib-Pb}.  The 
desired fractions are then given by:
\begin{equation}
f_{\rm{b}}^h \ = \ 
\frac{1}{P_{\rm{b}}} \frac{N_h^{ \rm b-tag}}{N_h} 
\left(\frac{2N_{\rm had}R_{\rm{b}}}
{N^{\rm b-tag}}\right)-\frac{1-P_{\rm{b}}}{P_{\rm{b}}}{\cal{B}}_h
\end{equation}
and are listed in Table~3.  Here $N^{ \rm b-tag}_h$ is the observed number 
of events with a tagged hadron $h$ in one hemisphere and a secondary b vertex 
in the one opposite and $N^{ \rm b-tag}$ is the total number of hemispheres 
with a secondary b vertex.  The term ${\cal{B}}_h$ takes into account the 
contribution to the measured fraction from events tagged as a b event but 
which do not come from $\rm Z^0\to b\overline{b}$ events.  This term, which 
is weighted by $(1-P_{\rm{b}})$, is calculated from Monte Carlo simulated 
events and is the only model-dependent number used in the measurement of 
the backgrounds from $\rm Z^0\to b\overline{b}$ events.  A conservative 
error of 50\% is assigned to the calculated value of ${\cal{B}}_h$, which
for pions, for example, is calculated to be ${\cal{B}}_{\pi^\pm}=0.127\pm 
0.015\pm 0.064$, where the first error is statistical and the second 
systematic. The values for the other hadron types are similar.   Given the 
very high purity of the b tagged sample, the influence of this term on the 
final measurement is small.  However, other sources of systematic uncertainty, 
such as uncertainties in $R_{\rm{b}}$ and $P_{\rm{b}}$, are negligible in 
comparison.  Due to low statistics, $f^{\Lambda(\overline{\Lambda})}_{\rm{b}}$ 
is taken from the simulation with a $\pm50\%$ relative systematic error, as 
given in Table~3.

In order to verify that the method does not introduce systematic biases, the 
measurement is repeated on JETSET events with a detailed simulation of the 
OPAL detector and the results are compared with the true b fractions in the 
Monte Carlo event sample.  
No systematic biases are found. The measurement is also checked
with the data by selecting b events using leptons with a high transverse 
momentum with respect to the jet direction (described in more detail below). 
The two measurements are found to be in good agreement, although the 
statistical errors for the lepton samples are large.  The dependence of the 
measurement on the purity of the b tagged sample is also checked by varying 
the minimum decay length significance over the range from 6 to 14. No 
systematic dependence of the measured b fractions is found.  
 
\begin{table}[tbh]
\renewcommand{\arraystretch}{1.2}
\begin{center}
\begin{tabular}{|c||c||c|} \hline
hadron type & $f_{\rm{b}}^h$ & $r_{\rm{b}}^h$         \\ \hline\hline
$\pi^\pm $            &  0.078$\pm$0.004 & 0.79$\pm$0.11  \\ \hline
$\rm K^\pm$           &  0.039$\pm$0.004 & 0.67$\pm$0.12  \\ \hline
$\rm p(\overline{p})$ &  0.051$\pm$0.009 & 0.71$\pm$0.30  \\ \hline
$\rm K^0_S$
                      &  0.036$\pm$0.010 &      ---       \\ \hline
$\Lambda(\overline{\Lambda})$ 
                      &  0.031$\pm$0.015 & 0.29$\pm$0.11  \\ \hline
\end{tabular}
\caption{Fractions and reliabilities of hadrons from bottom events,
where the errors include both statistical and systematic uncertainties
added in quadrature.}
\end{center}
\renewcommand{\arraystretch}{1.0}
\end{table}

To obtain the high-$x_p$ hadron charge reliabilities in bottom events one has 
to know the charge of the tagged bottom hadron. Instead of secondary vertices,
we therefore use high-energy muons to identify bottom particles. Details of
the muon identification and tagging procedures are given in~\cite{bib-muid}.
The observed charged tracks and neutral clusters in the electromagnetic 
calorimeter are combined into jets of a maximum observable invariant mass of 
7~GeV using the JADE algorithm with the `E0' recombination 
scheme~\cite{bib-JADEE0}. The event is tagged as a 
$\rm Z^0\to b\overline{b}$ event if a muon 
candidate is found with a total momentum greater than 3~GeV and a transverse 
momentum with respect to the jet axis, $p_T$, of at least 1~GeV.  

First we determine the overall charge reliability of the high-$p_T$ muon using 
double tagged b events. The b reliability is given by:
\begin{equation}
\left( 2 r_{\rm{b}} - 1 \right) \ = \
\sqrt{2\frac{N_{\rm bb}^{\rm OPP}}{N_{\rm bb}}-1} \ ,
\end{equation}
where $N_{\rm bb}$ is the total number of double tagged b events and 
$N_{\rm bb}^{\rm OPP}$ is the number of such events in which the 
tagging leptons have opposite charges.  We measure $r_{\rm{b}} = 0.650 
\pm 0.015$, where the error is statistical. Note that this measurement includes 
effects such as $\rm B^0 \overline{B}^0$ mixing. The high-$x_p$ hadron 
charge reliability is then given by:
\begin{equation}
\left( 2 r^h_{\rm{b}}-1 \right) \ = \ \frac {2 N_h^{ \rm b-tag,OPP} / 
N_h^{ \rm b-tag}-1} {2 r_{\rm{b}}-1 } \ ,
\label{eq-RELIAB}
\end{equation}
where $N_h^{ \rm b-tag,OPP}$ is the number of events in 
$N_h^{ \rm b-tag}$ in which the tagging particles have opposite charges. 
The results are also listed in Table~3.  Again, due to low statistics, 
$r^{\Lambda(\overline{\Lambda})}_{\rm{b}}$ is taken from the simulation with 
a systematic error corresponding to a relative error of $\pm 50\%$ on 
$(2 r^{\Lambda(\overline{\Lambda})}_{\rm{b}} -1)$, which is the factor 
relevant for the measurement of the forward-backward asymmetries.
 
\subsection{Charm Quark Contributions}

In the absence of an efficient enough charm tag we estimate the $f_{\rm{c}}^h$ 
from Monte Carlo simulations.  The estimated uncertainties on the 
$f_{\rm{c}}^h$ are based on the knowledge of 
charm production and decay.  The essential ingredients are the fragmentation 
functions of charm quarks~\cite{bib-CFRAGFCT} and the relative production 
yields of charmed hadrons~\cite{bib-CYIELDS} and their decay 
properties~\cite{bib-PDG}.  Taking into account the corresponding measurements 
and their uncertainties, we derive the fractions as listed in Table~4.
 
To cross-check these estimates with the data, we search in the 
tagged samples of high-$x_p$ particles for $\rm D^{*\pm}$ mesons 
in the opposite hemisphere.  We use the transition 
$\rm D^{*\pm }\rightarrow D^0\pi^\pm$ and the $\rm D^0$ decay modes
$\rm D^0\rightarrow K^-\pi ^+, \ K^-e^+\nu _e,$ and 
$\rm K^-\mu ^+ \nu _{\mu }$.  Details of the selection procedures are 
given in~\cite{bib-CASYMM}.  The results are consistent with the above 
estimates although with large statistical uncertainties. Results are reported 
for comparison in Table~4.

\begin{table}[tbh]
\renewcommand{\arraystretch}{1.2}
\begin{center}
\begin{tabular}{|c||c|c||c|} \hline
hadron type       
& $f_{\rm{c}}^h$ (JETSET)
% & $0.020\le\epsilon_{\rm c}\le 0.039$ 
& $f_{\rm{c}}^h$ (DATA)  
& $r_{\rm{c}}^h$ (JETSET)   \\ \hline
\hline
$\pi^\pm $        & 0.068$\pm$0.007 & 0.04 $\pm$0.04 
                  & 0.70$\pm$0.12  \\ \hline
$\rm K^\pm$       & 0.101$\pm$0.011 & 0.17 $\pm$0.06 
                  & 0.25$\pm$0.06  \\ \hline 
$\rm p(\overline{p})$ 
                  & 0.088$\pm$0.009 &                 
                  & 0.40$\pm$0.12  \\ \hline
$\rm K^0_S$   
                  & 0.114$\pm$0.012  &                 
                  &     ---        \\ \hline
$\Lambda(\overline{\Lambda})$ 
                  & 0.091$\pm$0.010 &                 
                  & 0.70$\pm$0.12  \\ \hline
\end{tabular}
\caption{Fractions and reliabilities for hadrons from charm events, where the 
errors include both statistical and uncorrelated systematic uncertainties added
in quadrature.  No measurements were possible for 
$f_{\rm{c}}^{\rm{p}(\overline{\rm p})}$, $f_{\rm{c}}^{\rm{K^0_S}}$, and 
$f_{\rm{c}}^{\Lambda(\overline{\Lambda})}$ in the data due to lack of 
statistics.  In addition to these uncertainties in the individual particle 
fractions, overall shifts between $-5\%$ and $+20\%$ are observed due to 
uncertainties in the charm fragmentation function and have to be taken into 
account.}
\end{center}
\renewcommand{\arraystretch}{1.0}
\end{table}

We also take the high-$x_p$ hadron charge reliabilities in charm events from 
studies of simulated events. The values are also given in Table~4. As a 
cross-check we measure the charge reliabilities from data using the 
$\rm D^{*^\pm}$ sample described above, using Eq.~\ref{eq-RELIAB}, but taking 
the overall charm reliability $r_{\rm{c}}$ from simulation. The results are 
compatible with the values from the simulation but have large statistical 
errors.

Several sources of systematic uncertainties have been considered.  Of 
particular relevance are those affecting the relative fractions of $\rm K^0_S$ 
and $\rm K^\pm$ from charm decays.  The ratio of $\rm K^0_S$ to $\rm K^\pm$ 
production depends on the inclusive branching ratios of the $\rm D^0$, 
$\rm D^\pm$, $\rm D_{\rm{s}}$ and $\Lambda^+_{\rm c}$ into the tagged 
particle species and the corresponding decay multiplicities.  The former 
were taken from~\cite{bib-PDG} and propagated through to the $f_{\rm c}^h$.  
The decay multiplicities as used in the JETSET simulation were compared to the 
measurements of the Mark~III collaboration~\cite{bib-Mark3}.
The differences lead to uncertainties of $\sim $4$\% $ for each particle
species.  The resulting errors for each of them are given in Table~4.  In 
addition, the $f_{\rm c}^h$ of all hadron species are affected coherently 
by the hardness of the fragmentation function.  Allowing the $\epsilon_{\rm c}$ 
parameter of the Peterson et al. fragmentation function~\cite{bib-Peterson} to 
vary between 0.020$\leq \epsilon_{\rm c} \leq $0.039, which corresponds to
$0.474\leq \langle x_E^{\rm D} \rangle \leq 0.501$, as given by the 
measurements of the $\rm{D}^*$ and $\rm D$ fragmentation 
functions~\cite{bib-CFRAGFCT}, overall changes of $-5\%$ to $+20\%$ are 
introduced. 
 
\section{The Branching Fractions and Forward-Backward \\
Asymmetries of Light Flavours}
 
As discussed in Section~2 the branching fractions of the Z$^0$ into light 
quarks and the forward-backward asymmetries are determined under the 
assumption that the electroweak couplings for down and strange quarks are
the same.  We start with the branching fractions and afterwards discuss the 
forward-backward asymmetries in Section~6.2, considering at first only 
statistical uncertainties.  The systematic uncertainties are addressed in 
Section~7.
 
\subsection{The Branching Fractions \boldmath $R'_q$}
 
After subtracting the contributions from heavy quarks, we solve the equation 
system for $R_{\rm{d,s}}$.  In addition to the five measured 
numbers of tagged hemispheres and fifteen combinations of double tagged events 
(Eqs.~1$-$2), we use the five hadronisation symmetries (Eqs.~4$-$8).  
Therefore, there are twenty measurements and twelve unknown parameters, 
including $R_{\rm{d,s}}$, a global $\rho$, and ten $\eta_q^h$.   
The equation system is solved by minimising a $\chi^2$ function which uses
the 20 measured quantities of Eqs.~1$-$2.   In order to minimise
the dependence of our results on heavy flavour electroweak measurements,
we determine the partial light flavour branching fractions:
\begin{equation}
R'_{\rm{d,s}} \ = \ 0.371 \pm 0.016,
\end{equation}
where the error is purely statistical and the value of $\chi^2$ is 4.4 for
eight degrees of freedom at the minimum.   The result for
$R'_{\rm{u}} \ = \ 0.258 \pm 0.031$ is fully anticorrelated to the 
value of $R'_{\rm{d,s}}$ since 
$R'_{\rm{d}}+R'_{\rm{u}}+R'_{\rm{s}}=1$.   The solution is tested for stability 
and uniqueness by using different starting values for the unknown parameters.  
No other physical solution is found.  The Standard Model 
expectations for a top quark mass of 175~GeV and a Higgs boson mass of 
300~GeV are 0.359 and 0.282, respectively, consistent with the above 
measurements.  If we use the world average values for $R_{\rm{c}}$ and 
$R_{\rm{b}}$~\cite{bib-PDG}, we can determine the branching fraction of the 
$\rm Z^0$ into down-type light quarks:
\begin{equation}
R_{\rm{d,s}} \ = \ 0.230 \pm 0.010 \ ,
\end{equation}
where again the error is purely statistical.  The corresponding 
$R_{\rm{u}}=0.160\pm 0.019$ is fully anticorrelated with this result. 

The values for the other parameters are given in Table~5, along with 
those obtained from the solution using Monte Carlo events for comparison.  
Some differences are found for $\eta_{\rm{d}}^{\pi^\pm}$ and for the baryons.  
The value of the hemisphere correlation, $\rho$, is comparable in data and 
Monte Carlo.  Most of the correlation is due to the limited geometric 
acceptance and the selection criteria, which are well modelled in the Monte 
Carlo simulations.  The correlation in JETSET events without detector 
simulation and without any restrictions on the geometric acceptance is about 
1.07 and is due to hemisphere correlations from gluon 
radiation~\cite{bib-letts+mattig}.

\begin{table}[tbh]
\renewcommand{\arraystretch}{1.2}
\begin{center}
\begin{tabular}{|c||c||cc|} \hline
parameter                               
&    OPAL                           & JETSET & (input)  \\ \hline\hline
$\rho$                                  
&  1.290  $\pm$ 0.043  $\pm$ 0.025  & 1.286  $\pm$ 0.032 & (1.238)  \\ \hline
$\eta_{\rm{d}}^{\pi^{\pm}}$         
&  0.0406 $\pm$ 0.0016 $\pm$ 0.0021 & 0.0441 $\pm$ 0.0012 & (0.0438) \\ \hline
$\eta_{\rm{s}}^{\pi^{\pm}}$         
&  0.0074 $\pm$ 0.0028 $\pm$ 0.0011 & 0.0084 $\pm$ 0.0021 & (0.0089) \\ \hline
$\eta_{\rm{d}}^{\rm K^{\pm}}$    
&  0.0010 $\pm$ 0.0052 $\pm$ 0.0012 & 0.0038 $\pm$ 0.0024 & (0.0045) \\ \hline
$\eta_{\rm{u}}^{\rm K^{\pm}}$   
&  0.0230 $\pm$ 0.0052 $\pm$ 0.0011 & 0.0178 $\pm$ 0.0025 & (0.0112) \\ \hline
$\eta_{\rm{s}}^{\rm K^{\pm}}$   
&  0.0344 $\pm$ 0.0029 $\pm$ 0.0013 & 0.0363 $\pm$ 0.0027 & (0.0407) \\ \hline
$\eta_{\rm{d}}^{\rm{p(\overline{\rm p})}}$        
&  0.0006 $\pm$ 0.0027 $\pm$ 0.0008 & 0.0065 $\pm$ 0.0032 & (0.0042) \\ \hline
$\eta_{\rm{u}}^{\rm{p(\overline{\rm p})}}$        
&  0.0082 $\pm$ 0.0056 $\pm$ 0.0015 & 0.0086 $\pm$ 0.0048 & (0.0124) \\ \hline
$\eta_{\rm{s}}^{\rm{p(\overline{\rm p})}}$        
&  0.0086 $\pm$ 0.0023 $\pm$ 0.0005 & 0.0042 $\pm$ 0.0017 & (0.0037) \\ \hline
$\eta_{\rm{d}}^{\Lambda(\overline{\Lambda})}$             
&  0.0049 $\pm$ 0.0020 $\pm$ 0.0004 & 0.0032 $\pm$ 0.0014 & (0.0022) \\ \hline
$\eta_{\rm{s}}^{\Lambda(\overline{\Lambda})}$             
&  0.0040 $\pm$ 0.0033 $\pm$ 0.0005 & 0.0107 $\pm$ 0.0024 & (0.0129) \\ \hline
\end{tabular}
\caption{Free parameters in the equation system and their fitted values,
with statistical and systematic errors, which are highly correlated.  
Also given for comparison are the values from a solution using JETSET events 
with full detector simulation with statistical errors only and the input
values given in parentheses.  Good agreement between the Monte Carlo input 
values and solutions is found.}
\end{center}
\renewcommand{\arraystretch}{1.0}
\end{table}

\subsection{The Forward-Backward Asymmetries}
 
To determine the forward-backward asymmetries, we divide the data sample
into events collected within $\pm 200$~MeV of the $\rm{Z}^0$ peak with an 
average centre-of-mass energy $\langle E_{\rm cm}\rangle =91.2$~GeV, and those 
below and above the peak, which have average centre-of-mass energies of 
$\langle E_{\rm cm}\rangle=89.5$~GeV and $\langle E_{\rm cm}\rangle=92.9$~GeV,
respectively.  The backgrounds due to charm and bottom events are known since 
the heavy flavour terms in Eq.~12 have been measured separately\footnote{See 
Tables~3-4 for the heavy flavour fractions $f_q^h$ and reliabilities $r_q^h$, 
and~\cite{bib-PDG} for $A_{FB}({\rm c})$ and $A_{FB}({\rm b})$.}.  We then use 
Eq.~12 together with the relation of Eq.~13 to determine the reliabilities of 
the charge tags for the light flavours.  The measured hadron forward-backward 
asymmetries, which have been determined by maximising a log-likelihood function,
are listed in Table~6.  
 
\begin{table}[tbh]
\renewcommand{\arraystretch}{1.2}
\begin{center}
\begin{tabular}{|c||c|c|c|c|} \hline
hadron type & 
$E_{\rm cm}<m_{\rm{Z}^0}$ & 
$E_{\rm cm}\sim m_{\rm{Z}^0}$ & 
$E_{\rm cm}>m_{\rm{Z}^0}$ \\ \hline\hline
$\pi^\pm$                     
& $(-1.22\pm2.14)\times 10^{-2}$ 
& $(-1.60\pm0.52)\times 10^{-2}$ 
& $(+0.12\pm1.78)\times 10^{-2}$ \\ \hline
$\rm K^\pm$                       
& $(+1.76\pm2.52)\times 10^{-2}$ 
& $(-3.12\pm0.58)\times 10^{-2}$ 
& $(-6.66\pm2.04)\times 10^{-2}$ \\ \hline
$\rm p(\overline{p})$             
& $(+9.98\pm5.18)\times 10^{-2}$ 
& $(-1.28\pm1.26)\times 10^{-2}$ 
& $(+3.42\pm4.42)\times 10^{-2}$ \\ \hline
$\Lambda(\overline{\Lambda})$ 
& $(+10.0\pm12.8)\times 10^{-2}$ 
& $(+4.32\pm3.18)\times 10^{-2}$ 
& $(+8.8\pm11.4)\times 10^{-2}$ \\ \hline
\end{tabular}
\caption{The measured forward-backward asymmetries, $A_{FB}^{h_{\rm det}}$, 
with statistical errors for various centre-of-mass energies, $E_{\rm cm}$.
Note that the $A_{FB}^{h_{\rm det}}$ have not been corrected for the purities 
of identified samples.  Results are given in percent.}
\end{center}
\renewcommand{\arraystretch}{1.0}
\end{table}

We extend the assumption $R_{\rm{d}}=R_{\rm{s}}$ to the equality of 
the forward-backward asymmetries of down and strange quarks, and use the 
hadronisation symmetries of Eqs.~14$-$16 and the constraint from Eq.~17. 
This leaves 8 unknown reliabilities, 10 combinations of double tagged events
of same and opposite charges (note that double tagged events with $K^0_S$ 
cannot be used), and 4 measured hadron $A_{FB}^{h_{\rm det}}$ with which to 
determine the light flavour forward-backward asymmetries.  We find for 
energies near the $\rm{Z}^0$ peak:
\begin{equation}
A_{FB}({\rm{d,s}}) \ = \ 0.068 \pm 0.035 
\ \ {\rm and} \ \ \ \ 
A_{FB}({\rm{u}})   \ = \ 0.040 \pm 0.067 \ ,
\end{equation}
where the errors given are statistical only.  The correlation between these 
results is +91\%.   The reliabilities are found to be consistent with the 
JETSET predictions, although with large statistical errors.  The three best 
measured reliabilities are $r_{\rm{d}}^{\pi^\pm}  =0.81 \pm 0.02$,
$r_{\rm{u}}^{\rm K^\pm}=0.70 \pm 0.18$, and 
$r_{\rm{s}}^{\rm K^\pm}=1.05 \pm 0.06$, where the errors are statistical.  
Near the $\rm Z^0$ peak, the flavour composition changes very slowly
and QCD scaling violations are likewise negligible. 
This is consistent with results from studies using JETSET where 
the flavour fractions and reliabilities for the off-peak data 
are found to be the same as for the on-peak data.  Therefore, we use the
flavour fractions and reliabilities as determined from the on-peak data to
determine the off-peak forward-backward asymmetries.  We find for energies 
above the $\rm{Z}^0$ peak:
\begin{equation}
A_{FB}({\rm d,s}) \ = \  0.228 \pm 0.082 
\ \ \ {\rm and} \ \ \                       
A_{FB}({\rm u})   \ = \ 0.397 \pm 0.183 , 
\end{equation}
and below the $\rm{Z}^0$ peak:
\begin{equation}
A_{FB}({\rm d,s}) \ = \ 0.019 \pm 0.098, 
\ \ \ {\rm and} \ \ \ 
A_{FB}({\rm u}) \ = \ 0.103 \pm 0.216 \ .
\end{equation}
The values as a function of the centre-of-mass energy are shown in Fig.~4,
along with the predictions of the Standard Model, which are found to be 
consistent with the data within the statistical errors.  The systematic 
errors are discussed in the following section.  

\section{Systematic Uncertainties}

The procedure used in this paper is tested with approximately eight million
hadronic $\rm{Z}^0$ decays generated with the JETSET model and passed through 
a detailed simulation of the OPAL detector.  The results obtained from the fit,
to the reconstructed events, $R_{\rm d,s}=0.224\pm 0.008$ and 
$A_{FB}({\rm d,s})= 0.128\pm0.025$, where the errors are statistical only, are 
found to be in agreement with the input values for $R_{\rm d,s}=0.220$ and 
$A_{FB}({\rm d,s})=0.100$.
 
Several sources of systematic uncertainty affect our result.  They can be 
divided into three classes: 1) those due to detector effects, 2) those due to 
model assumptions, and 3) uncertainties in the heavy quark contributions.  
Since these three classes of errors are largely uncorrelated, we estimate 
their individual impact on the branching fractions and asymmetries and add 
them quadratically to obtain the overall systematic error.  The individual 
systematic errors on the electroweak observables are summarised in Table~7 for 
$R_{\rm{d,s}}$ and in Table~8 for the forward-backward asymmetries.
The errors are determined by changing each input parameter in turn, repeating 
the analysis, and observing the shifts in the results, as given in the tables.

\begin{table}[p]
\renewcommand{\arraystretch}{1.30}
\begin{center}
\begin{tabular}{|l|c|c|} \hline
Source of Error   
& $\delta R'_{\rm{d,s}}$
& $\delta R_{\rm{d,s}}$ \\ \hline \hline
d$E$/d$x$ resolution $\pm 5\%$        & $\mp$0.0037 & $\mp$0.0023 \\ \hline
d$E$/d$x$ mean  $\pm 0.15\,\sigma$    & $\pm$0.0016 & $\pm$0.0010 \\ \hline
$\rm K^0_S$ purity and efficiency $\pm 4.1\%$ 
                                      & $\mp$0.0124  & $\mp$0.0077 \\ \hline
$\Lambda$ purity and efficiency $\pm 6.3\%$         
                                      & $\pm$0.0003 & $\pm$0.0002 \\ \hline
\hline
$\eta_{\rm{u}}^{\pi^\pm}/\eta_{\rm{d}}^{\pi^\pm}=1.00\pm0.02$ 
&  $<0.0001$  &  $<0.0001$  \\ \hline   
$\eta_{\rm{d}}^{\rm K^0(\overline{\rm K}^0)}/\eta_{\rm{u}}^{\rm K^\pm},\ 
 \eta_{\rm{u}}^{\rm K^0(\overline{\rm K}^0)}/\eta_{\rm{d}}^{\rm K^\pm}
=1.00\pm0.02$ 
& $\mp$0.0039 & $\mp$0.0024 \\ \hline   
$\eta_{\rm{s}}^{\rm K^0(\overline{\rm K}^0)}/\eta_{\rm{s}}^{\rm K^\pm}
=1.00\pm0.02$
& $\mp$0.0022 & $\mp$0.0014 \\ \hline   
$\eta_{\rm{u}}^{\Lambda(\overline{\Lambda})}
/\eta_{\rm{d}}^{\Lambda(\overline{\Lambda})}=1.00\pm0.02$          
& $\mp$0.0002 & $\mp$0.0001 \\ \hline   
$\rho$(baryon)$=\rho\times(1.00\pm0.02)$ 
& $\mp$0.0021 & $\mp$0.0013 \\ \hline \hline  
$f_{\rm{c}}^{\pi^\pm}=0.068\pm 0.007$   & $\mp$0.0005 & $\mp$0.0003 \\ \hline
$f_{\rm{c}}^{\rm K^\pm}=0.101\pm 0.011$ & $\pm$0.0050 & $\pm$0.0031 \\ \hline
$f_{\rm{c}}^{\rm p(\overline{p})}=0.088\pm 0.009$ 
                                        & $\mp$0.0011 & $\mp$0.0007 \\ \hline
$f_{\rm{c}}^{\rm K^0_S}=0.114\pm 0.012$   
                                        &  $\mp$0.0042 & $\mp$0.0026 \\ \hline
$f_{\rm{c}}^{\Lambda(\overline{\Lambda})}=0.091\pm 0.010$ 
                                        & $<0.0001$    & $<0.0001$    \\ \hline
$0.020 \le \epsilon_{\rm c} \le 0.039$  & $\pm$0.0015  & $\pm$0.0009 \\ \hline
\hline
$f_{\rm{b}}^{\pi^\pm}=0.078\pm 0.004$               
& $\mp$0.0003 & $\mp$0.0002 \\ \hline
$f_{\rm{b}}^{\rm K^\pm }=0.039\pm 0.004$               
& $\pm$0.0016 & $\pm$0.0010 \\ \hline
$f_{\rm{b}}^{\rm p(\overline{p})}=0.051\pm 0.009$               
& $\mp$0.0011 & $\mp$0.0007 \\ \hline
$f_{\rm{b}}^{\rm K^0_S}=0.036\pm 0.010$               
& $\mp$0.0033 & $\mp$0.0021 \\ \hline
$f_{\rm{b}}^{\Lambda(\overline{\Lambda})}=0.031\pm 0.015$               
& $\pm$0.0002 & $\pm$0.0001 \\ \hline \hline
Sub-total Systematic Error & 0.0160 & 0.0097 \\ \hline \hline
$R_{\rm{c}}=0.158 \pm0.010 $ & $<0.0001$ & $\mp$0.0037 \\ \hline
$R_{\rm{b}}=0.2212\pm0.0019$ & $<0.0001$ & $\mp$0.0008 \\ \hline\hline
Total Systematic Error       & 0.0160 & 0.0104 \\ \hline
Total Statistical Error      & 0.0155 & 0.0096 \\ \hline \hline
Total Error                  & 0.0223 & 0.0141 \\ \hline 
\end{tabular}
\caption{Sources of systematic error and their effects on the measurement
of $R'_{\rm{d,s}}$ and $R_{\rm{d,s}}$.  The corresponding errors on 
$R'_{\rm{u}}$ and $R_{\rm{u}}$ are twice those of $R'_{\rm{d,s}}$ and 
$R_{\rm{d,s}}$ and are negatively correlated, with the exception of the 
uncertainties due to $R_{\rm{c}}$ and $R_{\rm{b}}$, which are positively 
correlated.} 
\end{center}
\renewcommand{\arraystretch}{1.0}
\end{table}

\begin{table}[p]
\renewcommand{\arraystretch}{1.15}
\begin{center}
\begin{tabular}{|l|c|c|} \hline
Source of Error & $\delta A_{FB}({\rm{d,s}})$ 
& $\delta A_{FB}({\rm{u}})$ \\ \hline \hline 
d$E$/d$x$ resolution $\pm 5\%$         
& $\mp$0.0018  & $\mp$0.0032 \\ \hline
d$E$/d$x$ mean  $\pm 0.15\,\sigma$     
& $\pm$0.0009  & $\pm$0.0025 \\ \hline
$\rm K^0_S$ purity and efficiency $\pm 4.1\%$      
& $\mp$0.0051  & $\mp$0.0152 \\ \hline
$\Lambda$ purity and efficiency $\pm 6.3\%$      
& $\pm$0.0002  & $\pm$0.0006 \\ \hline\hline
$r_{\rm{d}}^{\rm K^\pm}=0.20\pm0.10$                    
& $\mp$0.0009  & $\mp$0.0013    \\ \hline
$r_{\rm{u}}^{\pi^\pm}/r_{\rm{d}}^{\pi^\pm}=1.00\pm0.05$          
& $\mp$0.0027  & $\mp$0.0072    \\ \hline
$r_{\rm{s}}^{\pi^\pm}=0.50\pm0.05$                  
& $\mp$0.0014  & $\mp$0.0010    \\ \hline
$r_{\rm{u}}^{\Lambda(\overline{\Lambda})}
/r_{\rm{d}}^{\Lambda(\overline{\Lambda})}=1.00\pm0.05$  
& $\mp$0.0013  & $\mp$0.0024    \\ \hline
$\rho$(baryon)$=\rho\times(1.00\pm0.02)$  
& $\mp$0.0032  & $\mp$0.0059    \\ \hline \hline  
$f_{\rm{c}}^{\pi^\pm}     =0.068\pm 0.007$          
& $\pm$0.0003  & $\mp$0.0012 \\ \hline
$f_{\rm{c}}^{\rm K^\pm}       =0.101\pm 0.011$          
& $\mp$0.0009  & $\pm$0.0020 \\ \hline
$f_{\rm{c}}^{\rm p(\overline{\rm p})}       =0.088\pm 0.009$          
& $\pm$0.0004  & $\mp$0.0003 \\ \hline
$f_{\rm{c}}^{\rm K^0_S} =0.114\pm 0.012$          
& $\mp$0.0012  & $\mp$0.0046 \\ \hline
$f_{\rm{c}}^{\Lambda(\overline{\Lambda})} =0.091\pm 0.010$          
&  $<0.0001$   & $\pm$0.0002 \\ \hline
$0.020 \le \epsilon_{\rm c} \le 0.039$ 
& $\mp$0.0022 & $\mp$0.0045 \\ \hline
$r_{\rm{c}}^{\pi^\pm}     =0.70 \pm 0.12$           
& $\mp$0.0012  & $\mp$0.0087 \\ \hline
$r_{\rm{c}}^{\rm K^\pm}       =0.25 \pm 0.06$           
& $\pm$0.0026  & $\pm$0.0044 \\ \hline
$r_{\rm{c}}^{\rm p(\overline{\rm p})}       =0.40 \pm 0.12$           
& $\mp$0.0018  & $\mp$0.0033 \\ \hline
$r_{\rm{c}}^{\Lambda(\overline{\Lambda})} =0.70 \pm 0.12$           
& $\mp$0.0002  & $\mp$0.0002 \\ \hline\hline
$f_{\rm{b}}^{\rm \pi^\pm}     =0.078\pm 0.004$          
& $\pm$0.0008  & $\pm$0.0020 \\ \hline
$f_{\rm{b}}^{\rm K^\pm}       =0.039\pm 0.004$          
& $\mp$0.0007  & $\pm$0.0001 \\ \hline
$f_{\rm{b}}^{\rm p(\overline{\rm p})}       =0.051\pm 0.009$          
& $\pm$0.0009  & $\pm$0.0007 \\ \hline
$f_{\rm{b}}^{\rm K^0_S} =0.036\pm 0.010$          
& $\mp$0.0009  & $\mp$0.0036 \\ \hline
$f_{\rm{b}}^{\Lambda(\overline{\Lambda})} =0.031\pm 0.015$          
& $\pm$0.0001  & $\pm$0.0003 \\ \hline 
$r_{\rm{b}}^{\rm \pi^\pm}= 0.79 \pm 0.11$          
& $\pm$0.0030  & $\pm$0.0129 \\ \hline
$r_{\rm{b}}^{\rm K^\pm}= 0.67 \pm 0.12$          
& $\mp$0.0039  & $\mp$0.0062 \\ \hline
$r_{\rm{b}}^{\rm p (\overline{\rm p})}= 0.71 \pm 0.30$          
& $\pm$0.0047  & $\pm$0.0078 \\ \hline 
$r_{\rm{b}}^{\Lambda(\overline{\Lambda})} = 0.29 \pm 0.11$          
&  $<0.0001$     &  $<0.0001$    \\ \hline \hline
$R_{\rm{c}}=0.158 \pm0.010 $                  
&  $<0.0001$   &  $<0.0001$  \\ \hline
$R_{\rm{b}}=0.2212\pm0.0019$                  
&  $<0.0001$     &  $<0.0001$    \\ \hline
$A_{FB}$(c)$ = 0.0722 \pm 0.0067$        
& $\mp$0.0011  & $\mp$0.0031 \\ \hline
$A_{FB}$(b)$ = 0.0992 \pm 0.0035$        
& $\pm$0.0001  & $\pm$0.0010 \\ \hline \hline
Total Systematic Error       & 0.0110       & 0.0281   \\ \hline
Total Statistical Error      & 0.0342       & 0.0667   \\ \hline\hline
Total Error                  & 0.0359       & 0.0723   \\ \hline 
\end{tabular}
\caption{Sources of systematic error and their effects on the 
measurements of the forward-backward asymmetries for 
$E_{\rm cm}\sim m_{\rm{Z}^0}$.}
\end{center}
\renewcommand{\arraystretch}{1.0}
\end{table}
 
\newpage

The following systematic uncertainties are considered:
\begin{itemize}
\item Detector sources:
   \begin{itemize}
      \item {\bf hadron identification by {\boldmath d$E$/d$x$}:}
            As mentioned in Section~3, we apply corrections to the apparent
            charged pion, charged kaon, and proton yields to take into account
            the migration of the true to the apparent particle species and
            their relative efficiencies.  The uncertainties in these 
            corrections are estimated by varying the widths, $\sigma$, of the 
            ionisation loss distributions by $\pm5\%$ and the central 
            values by $\pm 0.15\,\sigma$.  Since such changes affect all 
            charged particle species similarly, they result in smaller 
            uncertainties in the purities of the identified samples than
            the overall $\pi^\pm$, $\rm K^\pm$, and proton~(antiproton)
            identification efficiencies.
      \item {\bf efficiencies of {\boldmath$\rm K^0_S$} and 
            {\boldmath$\Lambda$}:}
            The limited knowledge of the $\rm K^0_S$ 
            and $\Lambda$ efficiencies 
            and purities affects the discrimination power between up and 
            down type quarks.  The $\rm K^0_S$ efficiency is one of the largest 
            sources of systematic uncertainty since the difference between 
            $\rm K^0_S$ and $\rm K^\pm$ production separates u and d quark 
            events, since $\rm K^0_S$ tags primarily d and s quarks while
            $\rm K^\pm$ tags primarily u and s quarks.    
      \item {\bf other selection criteria:} 
            To cross-check for possible biases, we repeat the analysis with 
            different maximum polar angles of the tagging particles between
            $0.6<|\cos\theta|<0.7$, different cuts on the polar angle of the 
            thrust axis $0.7<|\cos\theta_{\rm Thrust}|<0.9$, 
            and different minimum 
            values of $0.4<x_p<0.6$.  No significant deviations are observed.  

            Of the different minimum values of $x_p$ which were studied, 
            a minimum tagging hadron momentum $x_p >0.5$ was found to result
            in the smallest combined error.  The statistical error 
            grows with increasing $x_p$, while the heavy flavour backgrounds 
            (the uncertainty of which is a major source of systematic error) 
            decrease.  The measurements made using different minimum $x_p$ 
            give consistent results.
   \end{itemize}
\item Model uncertainties were addressed in some detail in~\cite{bib-letts+mattig}. 
      Here we summarise the important issues:
   \begin{itemize}
      \item {\bf hadronisation symmetries (Eqs.~4{\boldmath $-$}8):}
            The hadronisation relations of Eqs.~4$-$8 are stable against 
            variations of fragmentation parameters within 2\%.  As shown in 
            Table~7, the corresponding uncertainties in the branching 
            fractions are small.

            To estimate the uncertainties in the hadronisation relations,
            we vary several of the important fragmentation parameters in 
            JETSET\footnote{An alternative hadronisation model is implemented
            in the HERWIG~\cite{bib-HERWIG} generator.  However, this model is
            in general not SU(2) isospin symmetric due to technical 
            reasons~\cite{bib-technical}.
            Therefore, HERWIG
            is not suited for a general study of these hadronisation 
            relations.}. The default values are obtained from a fit to
            event properties such as charged particle multiplicity, the 
            $x_p$ distributions for $\pi^\pm$, $\rm K^\pm$, and protons, 
            particle production rates, thrust (and other event shape) 
            distributions~\cite{bib-toon}, etc.  Uncertainties in the 
            parameters can be estimated from the corresponding increase of 
            the $\chi^2$ of the model with respect to the data.  Requiring a 
            change of at least $\Delta \chi^2$=1, we vary the following 
            parameters:
            \begin{itemize}
               \item the hardness of the fragmentation, achieved by varying 
                     the `b' parameter of the LUND symmetric fragmentation
                     function~\cite{bib-LUND} between 0.22 and 0.62,
               \item the fraction of strange quarks in the sea, by varying 
                     the relevant JETSET parameter $\gamma_{\rm{s}}$ 
                     between 0.27 and 0.36,
               \item the fraction of vector mesons, varied between
                     0.4 and 0.7 for hadrons with only up and down
                     quarks and between 0.2 and 0.6 for those containing 
                     strange quarks,
               \item the production rates of tensor mesons by varying
                     {\tt PARJ(11)} between 0.4 and 0.7, {\tt PARJ(12)} between
                     0.2 and 0.6, and switching on tensor meson production,
               \item the value of $\Lambda_{\rm QCD}$ used in JETSET between
                     0.25 and 0.29.
            \end{itemize}
            The hadronisation symmetries show no statistically significant 
            variations outside the $\pm 2\%$ range.

            The relations are also confirmed by a study with the COJETS 
            model~\cite{bib-COJETS} which invokes independent fragmentation.
   
      \item {\bf hemisphere correlations:}
            Dynamic correlations and geometrical effects are taken into 
            account by the $\rho$ parameter, which is found to be independent 
            of the primary quark flavour and the tagging hadron type to good 
            accuracy in JETSET.  To study possible breaking of this 
            universality, we allow $\rho$ for the proton and the $\Lambda$ to 
            vary by $\pm 2\%$ relative to the overall $\rho$, thus allowing
            for possible differences between baryon and meson hadronisation.  
   \end{itemize}
\item Heavy flavour contributions:
  \begin{itemize}
      \item {\bf heavy flavour fractions and reliabilities:} 
            The estimation of the heavy flavour contributions was discussed in
            Section~4.  The largest uncertainties are due to the relative 
            amounts of charged and neutral kaons from charm events.  The 
            measured $\rm D^0$ and ${\rm D^{\pm}\rightarrow K^0_S}X$ vs.  
            ${\rm K^\pm}X$ branching fractions cause the largest systematic 
            uncertainties, as can be seen in Tables~7 and~8.  It is again the 
            uncertainty in the ability to separate up and down quarks with 
            charged and neutral kaons which limits the accuracy of the 
            measurements.
      \item {\bf heavy flavour Standard Model quantities:}
            Variations of the heavy flavour branching fractions of the 
            $\rm{Z}^0$ and the forward-backward asymmetries are studied and 
            found to have only small effects on the measurements of the light 
            flavour electroweak parameters (see Tables~7 and~8).  Note that
            the uncertainties on $R_{\rm{c}}$ and $R_{\rm{b}}$ have a 
            negligible effect on the measurements of $R'_q$.
   \end{itemize}
\end{itemize}

\section{Results and Conclusions}

Taking into account the systematic uncertainties, the final results 
(assuming $R'_{\rm{d}}=R'_{\rm{s}}$) are:
\begin{eqnarray}
R_{\rm{d,s}}' & \ = \ & 
0.371 \ \pm 0.016 \ \rm{(stat.)} \  
        \pm 0.016 \ \rm{(syst.)} \ \  \ \ \rm{and} \\ 
R'_{\rm{u}}   & \ = \ & 
0.258 \ \pm 0.031 \ \rm{(stat.)} \  
        \pm 0.032 \ \rm{(syst.)} \ . 
\end{eqnarray}
The Standard Model predictions are 0.359 and 0.282, respectively.  The $R'_q$ 
have the advantage of being largely independent of $R_{\rm{c}}$ and 
$R_{\rm{b}}$.  If we assume the world average values of $R_{\rm{c}}$ and 
$R_{\rm{b}}$ from~\cite{bib-PDG}, we find:
\begin{eqnarray}
R_{\rm{d,s}} & \ = \ & 
0.230 \ \pm 0.010 \ \rm{(stat.)} \  
        \pm 0.010 \ \rm{(syst.)} \ \  \ \ \rm{and} \\
R_{\rm{u}} & \ = \ & 
0.160 \ \pm 0.019 \ \rm{(stat.)} \  
        \pm 0.019 \ \rm{(syst.)} \ . 
\end{eqnarray}

For the forward-backward asymmetries, we determine:
\begin{eqnarray}
A_{FB}({\rm{d,s}}) & \ = \ & 
0.068 \ \pm 0.035 \ \rm{(stat.)} \  
        \pm 0.011 \ \rm{(syst.)} \ \  \ \ \rm{and} \\
A_{FB}({\rm{u}}) & \ = \ & 
0.040 \ \pm 0.067 \ \rm{(stat.)} \  
        \pm 0.028 \ \rm{(syst.)} 
\end{eqnarray}
for centre-of-mass energies near the $\rm{Z}^0$ peak.  The measurement of 
$A_{FB}({\rm{d,s}})$ is consistent with the result obtained by the DELPHI
collaboration~\cite{bib-DELFISTR} using a different method which is more 
dependent on hadronisation models.  Note that the measurements 
$R'_{\rm{d,s}}$ and $R'_{\rm{u}}$ are completely negatively 
correlated and those of $A_{FB}({\rm{d,s}})$ and 
$A_{FB}({\rm{u}})$ more than 90\% positively correlated.  All of the 
measurements are in agreement with the predictions of the Standard Model.

We interpret these results in terms of right and left handed 
${\rm{Z^0}}q\overline{q}$ couplings
$g_R^{\rm d,s}$ and $g_L^{\rm d,s}$ of down and strange quarks.  Here
$g_L^q = \rho_q(I_3^q - e_q\sin ^2\theta_{\rm eff})$ and
$g_R^q = -\rho_q e_q\sin^2\theta_{\rm eff}$, $\rho_q$ and 
$\sin^2\theta_{\rm eff}$ being the effective electroweak parameters, $I_3$
the third component of the weak isospin, and $e_q$ the quark electric charge
in units of the magnitude of the electron charge.  The left and right handed 
couplings are obtained by solving the equations
\begin{equation}
R'_{\rm d,s} = \frac{(g^{\rm d,s}_R)^2+(g^{\rm d,s}_L)^2}
{2[(g^{\rm d,s}_R)^2+(g^{\rm d,s}_L)^2]+(g^{\rm u}_R)^2+(g^{\rm u}_L)^2}
\end{equation}
\begin{equation}
A_{FB}({\rm{d,s}})(m_{\rm Z^0}) =
\frac{3}{4}{\cal A}_{\rm e}
\frac{(g^{\rm d,s}_R)^2-(g^{\rm d,s}_L)^2}{(g^{\rm d,s}_R)^2+(g^{\rm d,s}_L)^2}
\end{equation}
with the electron coupling ${\cal A}_{\rm e}=0.1466\pm 0.0033$~\cite{bib-PDG}.
The right and left handed couplings of up quarks are assumed to be
fully anticorrelated to the down/strange ones such that the denominator of
Eq.~35 is equal to the Standard Model expectation.  In addition, the energy
dependence of the off-peak asymmetries to lowest order was parametrised 
according to~\cite{bib-YB1}:
\begin{equation}
A_{FB}(s) = \frac{3}{4} \frac{G_3(s)}{G_1(s)}
\end{equation}
\begin{equation}
 G_1(s) \ = \ Q_{\rm e}^2 Q_f^2\, \mid \chi_{\gamma} \mid ^2  +
         2\, Q_{\rm e} Q_f\, v_{\rm e} v_f\, 
         {\rm Re}\, ( \chi_{\gamma}^* \chi_{\rm Z^0} ) 
        + (v_{\rm e}^2+a_{\rm e}^2)\, (v_f^2+a_f^2)\, 
        \mid \chi_{\rm Z^0} \mid^2  
\end{equation}
\begin{equation}
 G_3(s) \ = \ 2\, Q_{\rm e} Q_f\, a_{\rm e} a_f\, 
           {\rm Re}\, (\chi_{\gamma}^* \chi_{\rm Z^0})       
           + 4\, v_{\rm e} a_{\rm e} \, v_f a_f\, \mid \chi_{\rm Z^0} \mid^2    
\end{equation}
with
\begin{equation}
   \chi_{\rm Z^0} = 
   \frac{s}{s-m_{\rm Z^0}^2 + \imath \, \Gamma_{\rm Z^0}  s/m_{\rm Z^0} }
\end{equation}
and
\begin{equation}
 \chi_{\gamma} \ = \ \frac{1}{1+\Pi_{\gamma} } \ \ \ , \ \ \ \
  \Pi_{\gamma} \ = \ - 0.0593 \pm 0.0007. 
\end{equation}
The axial and vector couplings are related to the left and right handed 
couplings through $a_f=g_L^f-g_R^f$ and $v_f=g_L^f+g_R^f$.  The  values 
of the axial and vector couplings for the electron are taken 
from~\cite{bib-PDG}.

The measurements are corrected for initial photon radiation and QCD effects
using the JETSET model~\cite{bib-JETSET} including initial state photon 
radiation.  Both hard gluon and photon emission are suppressed due to the 
requirement of a high-$x_p$ particle.  The corrections are negligible for the 
measurements of $R_q$ and for the asymmetries amount to $+0.002$ due to QCD 
effects and $+0.004$, $+0.002$ and $+0.017$ due to initial state photon 
emission at centre-of-mass energies of 89.6, 91.2 and 92.9~GeV, respectively,
which are small compared to the statistical error.

We find $g^{\rm d,s}_L=-0.44^{+0.13}_{-0.09}$ and 
$g^{\rm d,s}_R=0.13^{+0.15}_{-0.17}$ in agreement with the Standard Model 
values of $-0.424$ and $+0.077$, respectively.
The results are shown in Fig.~5 together with various confidence level 
regions.   The fit yields a $\chi^2$ of 2.6 for two degrees of freedom.
The couplings enter quadratically into the on-peak observables but
linearly in the $\gamma$-$\rm Z^0$ interference contribution for the off-peak 
asymmetries. Therefore, only the latter measurements allow discrimination
between the possible signs of $g_L$ and $g_R$.   The solution for a 
negative $g_R$ leads to an acceptable $\chi ^2$=3.0 and cannot be excluded.
Solutions with a positive $g_L$ are disfavoured and can be excluded 
with 76\% and 83\% confidence levels for negative and positive $g_R$, 
respectively.  

Whereas results on down quarks exist from lepton-nucleon scattering, the 
present analysis is the most direct measurement of the strange quark couplings.
Departing from the assumption of the equality of the couplings of down
and strange quarks, we fix $R'_{\rm d}$ and determine $R'_{\rm s}$,
yielding the results presented in Fig.~6.  Fixing $R'_{\rm d}$ to the 
Standard Model value results in $R'_{\rm s}=0.392 \pm 0.043 \thinspace{\rm 
(stat.)} \pm 0.045 \thinspace{\rm (syst.)}$.   The dependence of
$R'_{\rm s}$ on $R'_{\rm d}$ is linear and can be parametrised as 
${\rm d}R'_{\rm s}/{\rm d}R'_{\rm d}=-1.83$.  Note that the value of 
$R'_{\rm u}$ is completely anticorrelated to $R'_{\rm s}$.
Similar results can be obtained for $A_{FB}({\rm s})$ and $A_{FB}({\rm u})$ 
for various fixed values of $A_{FB}({\rm d})$ as shown in Fig.~7.   
For $A_{FB}({\rm d})$ fixed to the Standard Model value of $0.100$, 
we find $A_{FB}({\rm s})= 0.075 \pm 0.028 \thinspace{\rm (stat.)} \pm 0.008
\thinspace{\rm (syst.)}$ and $A_{FB}({\rm u})=0.086 \pm 0.030
\thinspace{\rm (stat.)} \pm 0.021 \thinspace{\rm (syst.)})\%$.  In this case, 
$A_{FB}({\rm s})$ and $A_{FB}({\rm u})$ are only $+31\%$ correlated.  
Again, the dependence of the results on $A_{FB}({\rm d})$ is linear and can 
be parametrised by ${\rm d}A_{FB}({\rm s})/{\rm d}A_{FB}({\rm d})=+0.32$ and
${\rm d}A_{FB}({\rm u})/{\rm d}A_{FB}({\rm d})=+1.42$.  The statistical and 
systematic errors on $A_{FB}({\rm u})$ are independent of $A_{FB}({\rm d})$, 
while those of $A_{FB}({\rm s})$ scale linearly with the value of 
$A_{FB}({\rm s})$ such that the relative error is constant.

A measurement of all of the individual light-flavour electroweak parameters 
would require that the number of unknown flavour tagging efficiencies, 
$\eta_q^h$, be reduced.  This might be possible in the future by using 
additional hadronisation symmetries~\cite{bib-letts+mattig} which, however, may 
introduce a greater model dependence.   

\vfill

Acknowledgements:
\par
We particularly wish to thank the SL Division for the efficient operation
of the LEP accelerator and for their continuing close cooperation with
our experimental group. We thank our colleagues from CEA, DAPNIA/SPP,
CE-Saclay for their efforts over the years on the time-of-flight and trigger
systems which we continue to use.  In addition to the support staff at our own
institutions we are pleased to acknowledge the  \\
Department of Energy, USA, \\
National Science Foundation, USA, \\
Particle Physics and Astronomy Research Council, UK, \\
Natural Sciences and Engineering Research Council, Canada, \\
Israel Science Foundation, administered by the Israel
Academy of Science and Humanities, \\
Minerva Gesellschaft, \\
Benoziyo Center for High Energy Physics,\\
Japanese Ministry of Education, Science and Culture (the
Monbusho) and a grant under the Monbusho International
Science Research Program,\\
German Israeli Bi-national Science Foundation (GIF), \\
Direction des Sciences de la Mati\`ere du Commissariat \`a l'Energie
Atomique, France, \\
Bundesministerium f\"ur Bildung, Wissenschaft,
Forschung und Technologie, Germany, \\
National Research Council of Canada, \\
Hungarian Foundation for Scientific Research, OTKA T-016660,
T023793 and OTKA F-023259.\\

%%%%%%%%%%%%%%%%%%%%%%%%%%%%%%%%%%%%%%%%%%%%%%%%%%%%%%%%%%%%%%%%%%%

\newpage
\begin{figure}[p]
\epsfig{file=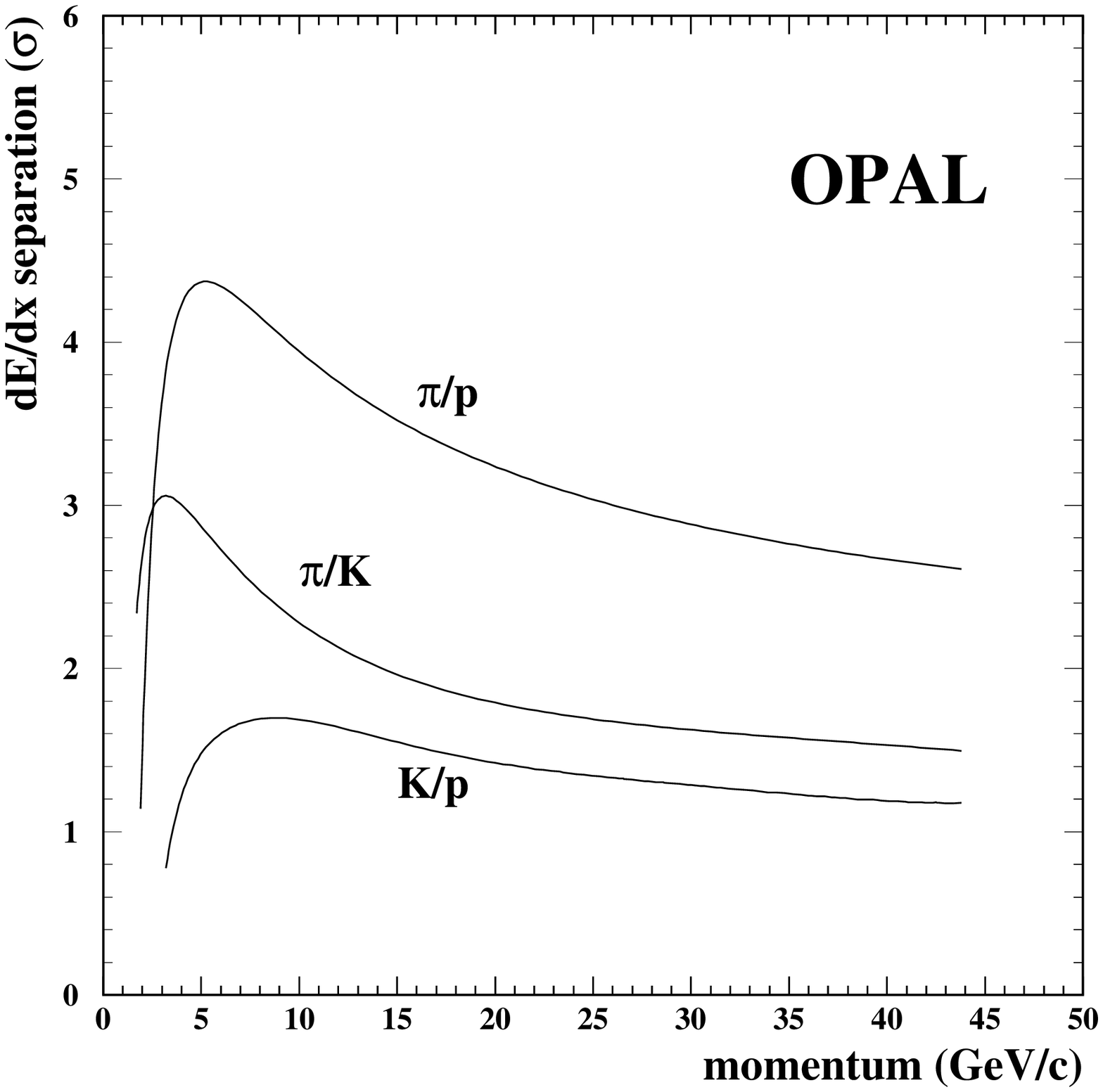,height=17cm}
\caption{Two particle separation power 
$({\rm d}E/{\rm d}x_1-{\rm d}E/{\rm d}x_2)/ 
     \langle\sigma({\rm d}E/{\rm d}x)\rangle$
in units of mean ${\rm d}E/{\rm d}x$ resolution
as a function of particle momentum for tracks with $|\cos\theta|<0.7$
and at least 20 measured samples, obtained from data.  The curves given 
are for pion/proton, pion/kaon and kaon/proton separation.
} 
\end{figure}

\newpage
\begin{figure}[p]
\epsfig{file=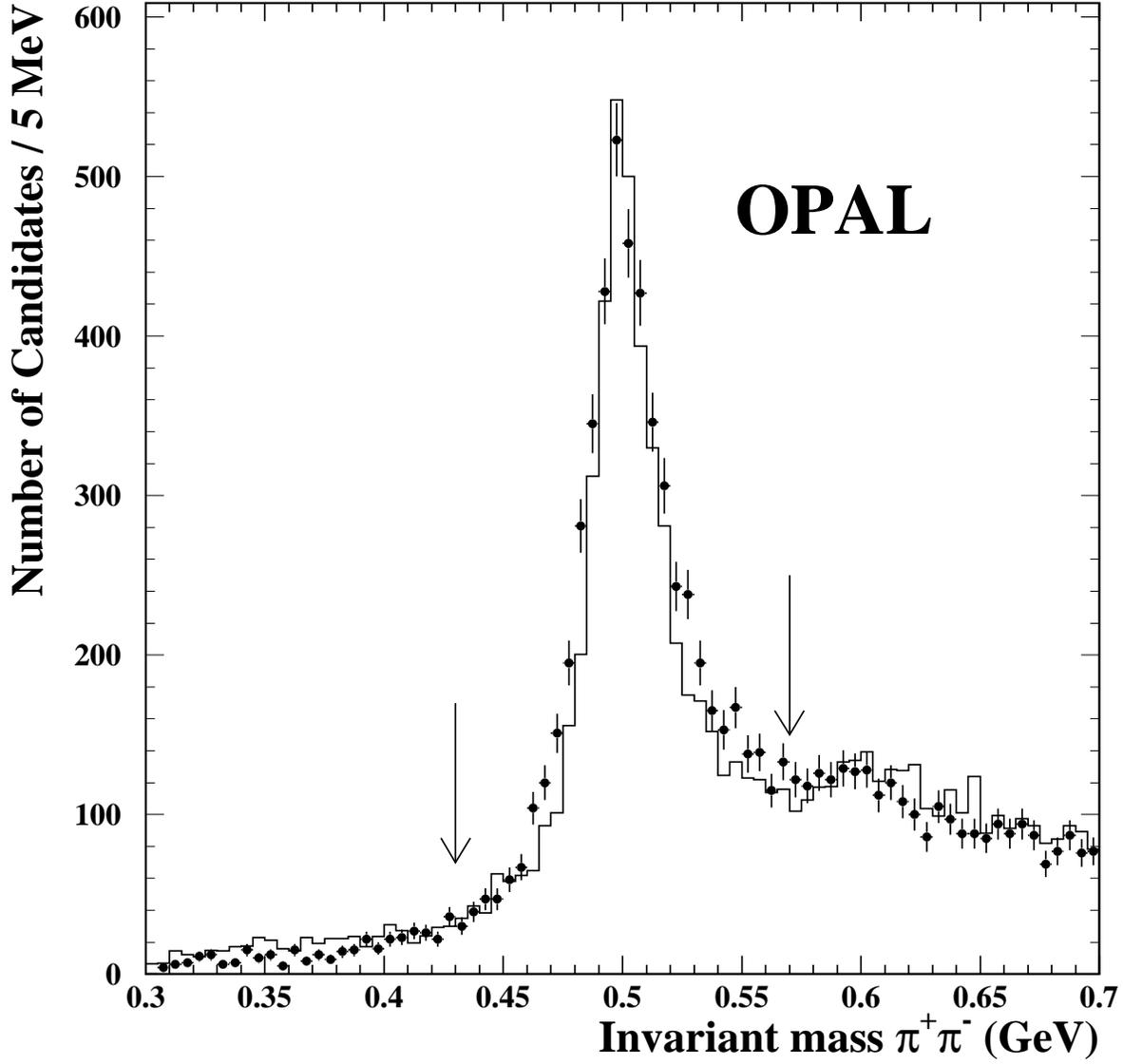,height=17cm}
\caption{Invariant mass $\pi^+\pi^-$ of $\rm K^0_S$ candidates with $x_p>0.5$ 
in the data (points with error bars) compared to Monte Carlo (histogram) which
is normalised to the same number of $\rm{Z}^0$ events.  Note that the details 
of the total $\rm K^0_S$ production rate in the Monte Carlo are not relevant 
to this analysis.  The arrows indicate the mass range in which $\rm K^0_S$ 
candidates are selected as high-$x_p$ tags.  The different mass resolutions
in the data and Monte Carlo are taken into account in the efficiency 
determination.}
\end{figure}

\newpage
\begin{figure}[p]
\epsfig{file=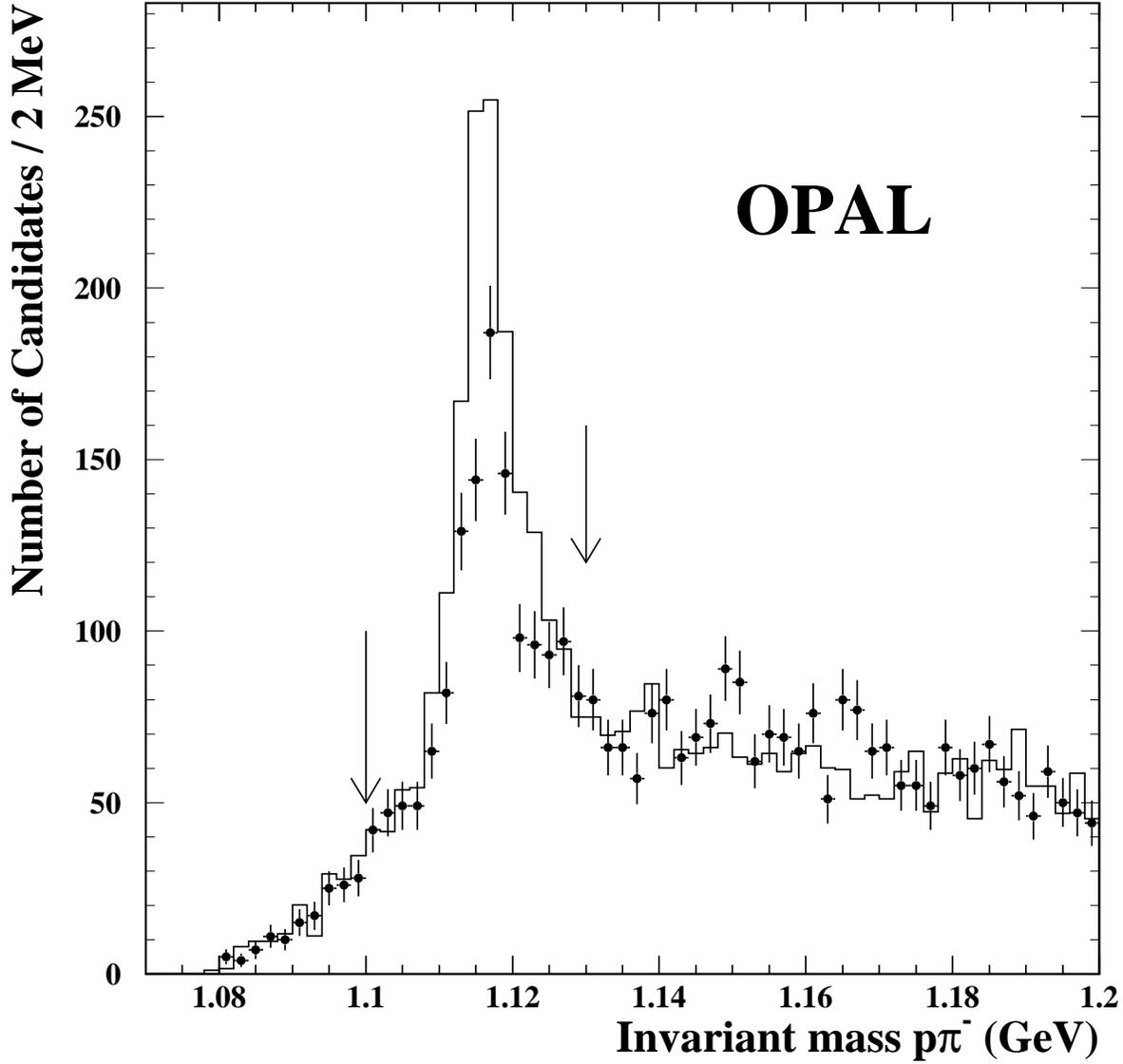,height=17cm}
\caption{Invariant mass ${\rm{p}}\pi^-$ of $\Lambda$ candidates 
with $x_p>0.5$ in the 
data (points with error bars) compared to Monte Carlo (histogram) which
has been normalised to the same number of $\rm{Z}^0$ events.  The arrows 
indicate the mass range in which $\Lambda$ candidates are selected as 
high-$x_p$ tags.  Note that the $\Lambda$ fragmentation function is too hard 
in JETSET, resulting in an overestimation of the high-$x_p$ production of 
$\Lambda$ baryons, and that the details of the total $\Lambda$ production 
rate in the Monte Carlo are not relevant to this analysis.}
\end{figure}

\newpage
\begin{figure}[p]
\epsfig{file=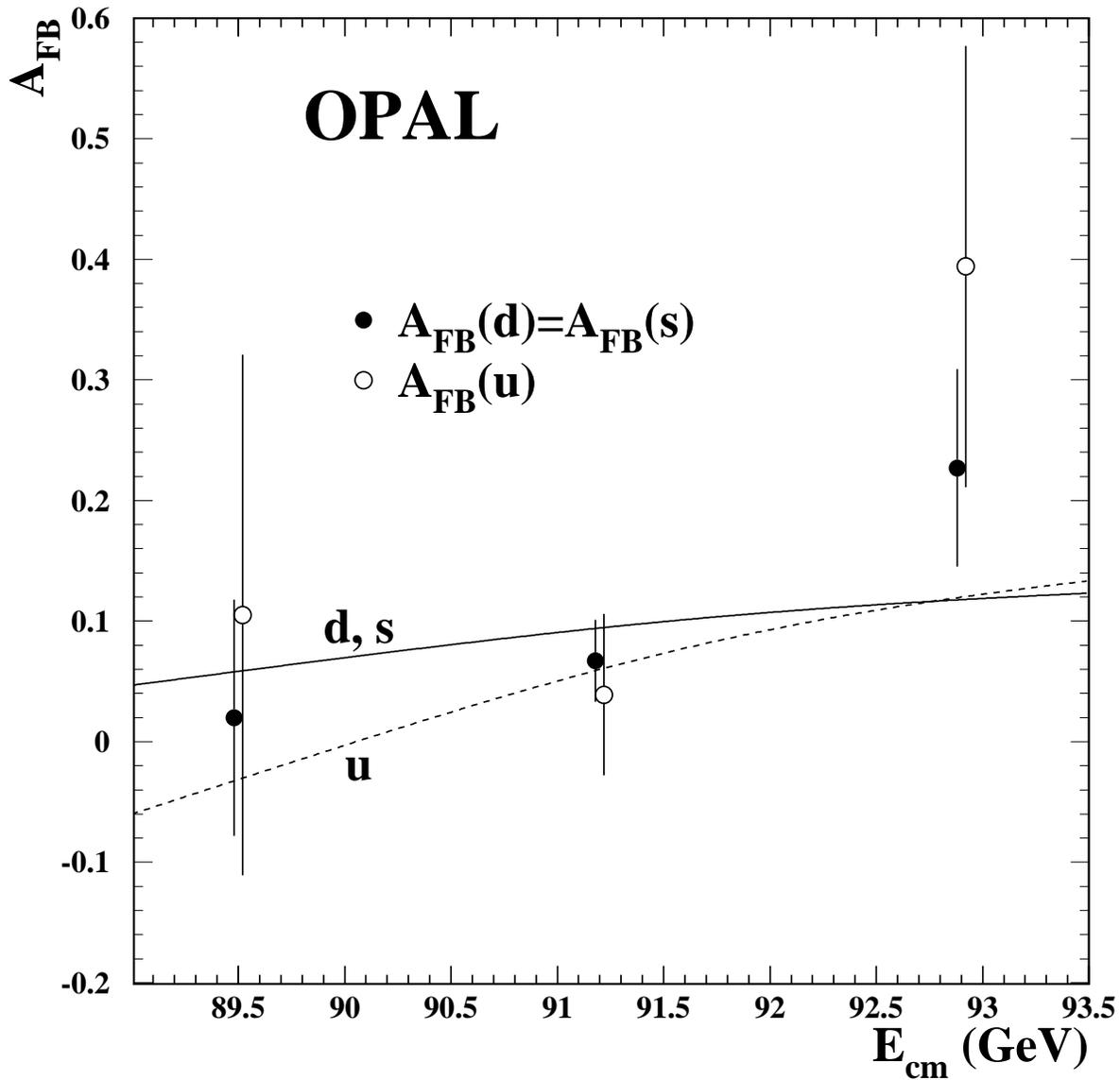,height=17cm}
\caption{$A_{FB}({\rm{d,s}})$ and $A_{FB}({\rm{u}})$ versus the centre-of-mass 
energy, $E_{\rm cm}$, where the 
errors are statistical plus systematic.  The statistical errors for different 
centre-of-mass energies are uncorrelated.  The systematic errors are correlated 
between centre-of-mass points.  The Standard Model 
predictions for $m_{\rm top}=175$~GeV and $m_{\rm Higgs}=300$~GeV are shown 
as the solid curve for $A_{FB}({\rm{d,s}})$ and the dashed 
curve for $A_{FB}({\rm{u}})$.}
\end{figure}

\newpage
\begin{figure}[p]
\epsfig{file=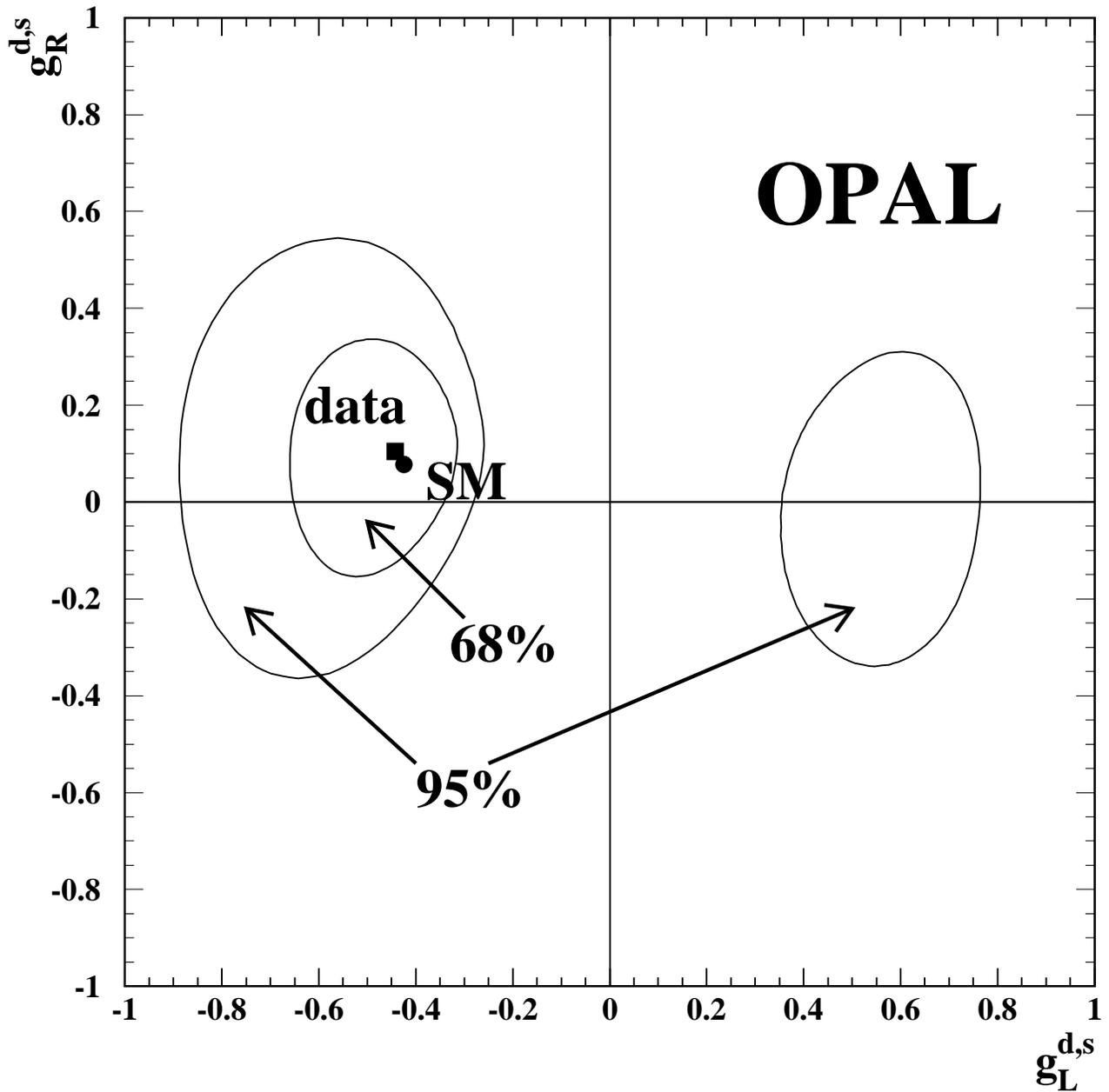,height=17cm}
\caption{Left vs. right handed couplings of strange quarks, assuming
identical values for down quarks.  The 68\% and 95\% confidence level
regions are also shown.  The points correspond to the measurement and 
the Standard Model expectation.}
\end{figure}

\newpage
\begin{figure}[p]
\epsfig{file=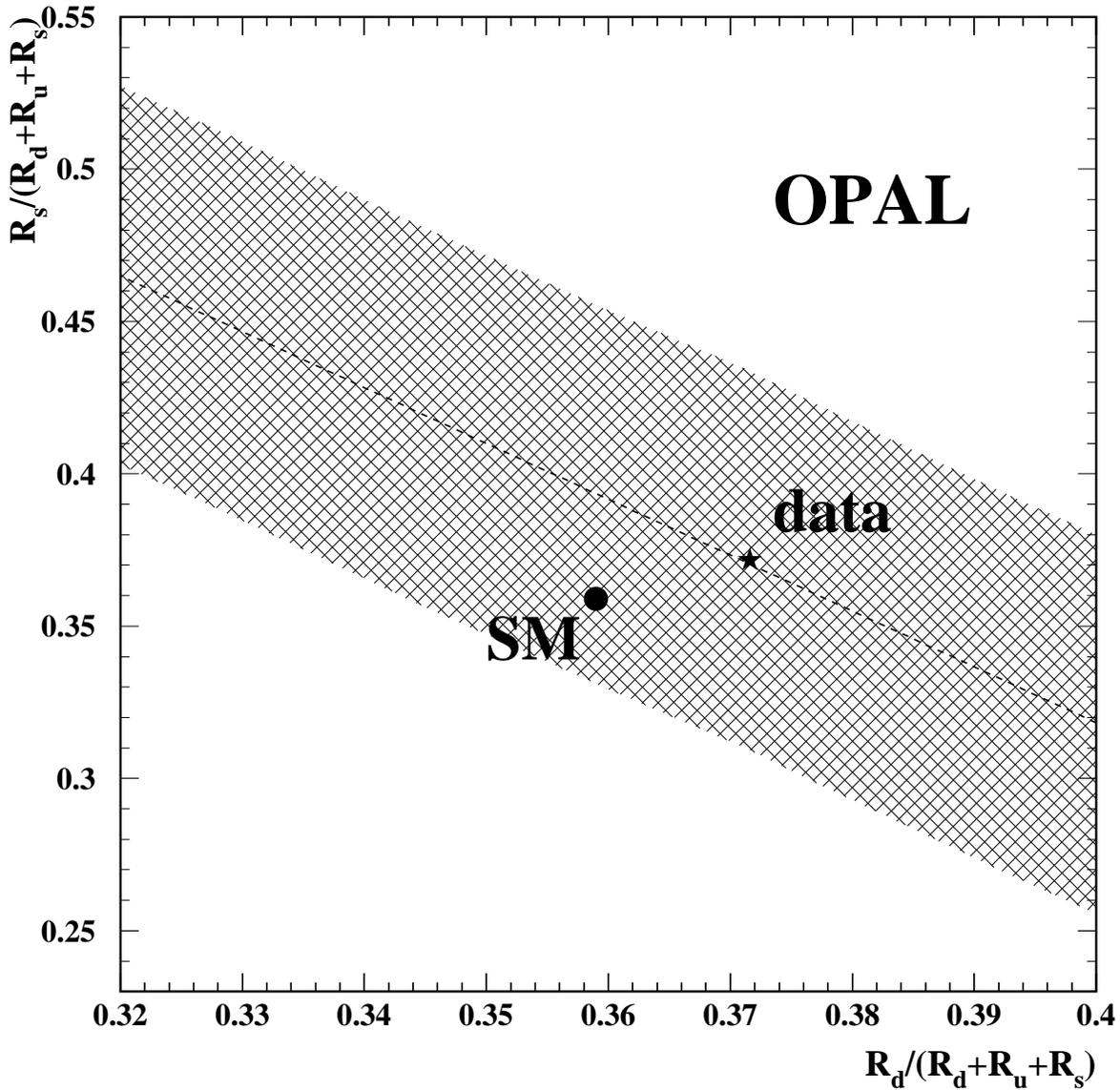,height=17cm}
\caption{
$R'_{\rm s}=R_{\rm s}/(R_{\rm d}+R_{\rm u}+R_{\rm s})$ 
determined for various fixed values of 
$R'_{\rm d}=R_{\rm d}/(R_{\rm d}+R_{\rm u}+R_{\rm s})$.
The result obtained by enforcing the equality of $R'_{\rm d}$ and
$R'_{\rm s}$ is given as the star, and the Standard Model value is shown 
as the solid point.  The hatched region shows the total error on $R'_{\rm s}$
for a given $R'_{\rm d}$.  Note that $R'_{\rm u}=1-R'_{\rm d}-R'_{\rm s}$.}
\end{figure}

\newpage
\begin{figure}[p]
\epsfig{file=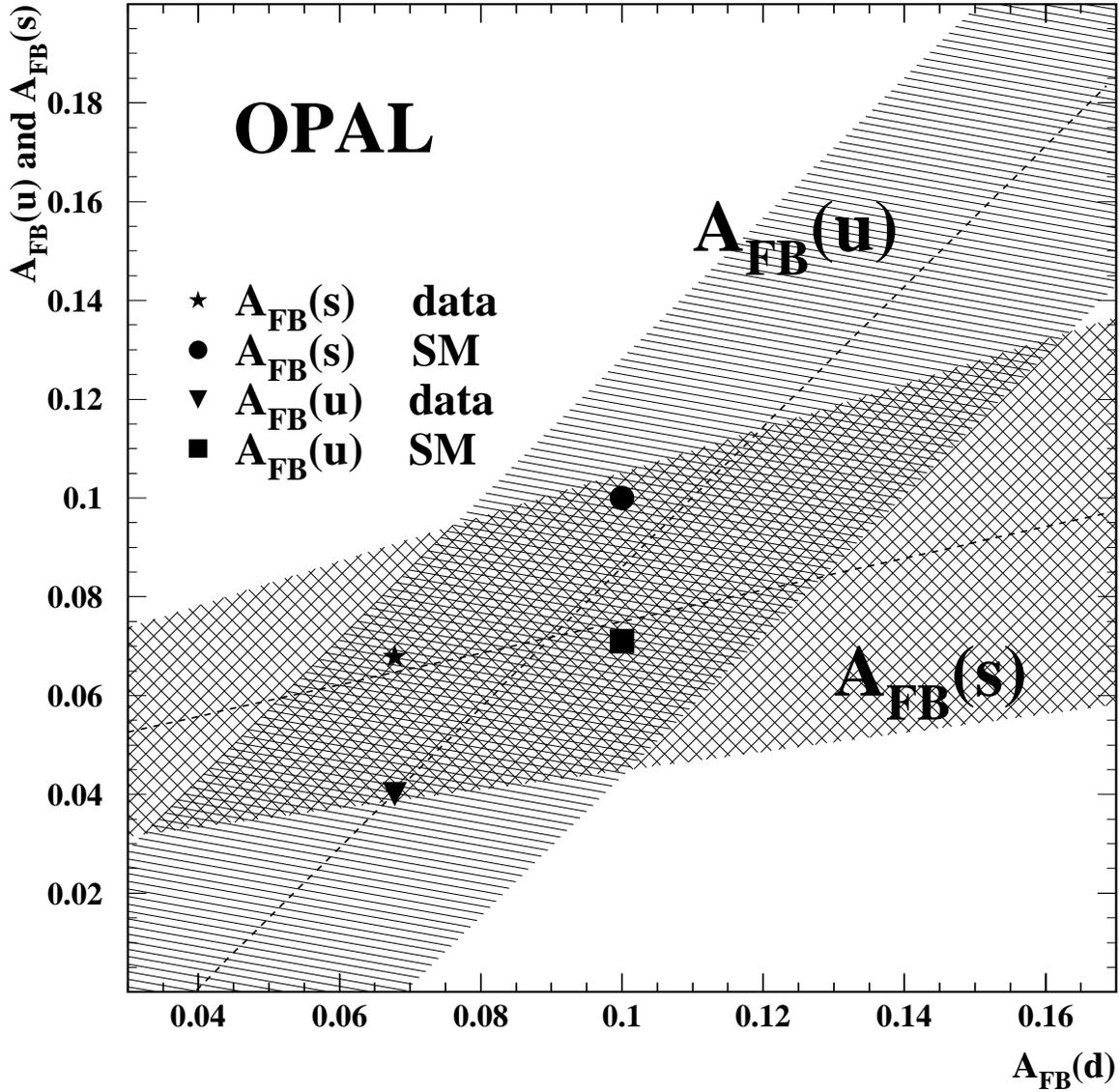,height=17cm}
\caption{$A_{FB}({\rm s})$ and $A_{FB}({\rm u})$ determined for various 
fixed values of $A_{FB}({\rm d})$.  The results obtained by enforcing the 
equality of the down-type light flavour couplings are shown as the triangle 
and the star and the Standard Model values are shown as the square and circle 
for u and s quarks, respectively.  The single and double hatched regions show 
the total error on $A_{FB}({\rm u})$ and $A_{FB}({\rm s})$, respectively,
for a given $A_{FB}({\rm d})$.}
\end{figure}

\end{document}